\documentclass[useAMS,usenatbib]{mnras}
\usepackage{graphicx,natbib,amssymb,amsmath,stmaryrd,makecell}
\usepackage{txfonts}
\usepackage{xcolor}
\usepackage{hyperref}
\usepackage{xspace}

\newcommand{\expf}[1]{{{\rm e}^{#1}}}

\newcommand{\beq}{\begin{equation}}   %

\newcommand{\eeq}{\end{equation}}   %

\newcommand{\beqa}{\begin{eqnarray}}   %

\newcommand{\eeqa}{\end{eqnarray}}   %

\newcommand{\beal}{\begin{align}}

\newcommand{\enal}{\end{align}}

\newcommand{\bspl}{\begin{split}}

\newcommand{\espl}{\end{split}}

\newcommand{\bsub}{\begin{subequations}}

\newcommand{\esub}{\end{subequations}}

\newcommand{\bmulti}{\begin{multline}}   %

\newcommand{\beqm}{\begin{mathletters}}   %

\newcommand{\eeqm}{\end{mathletters}}   %

\newcommand{\kB}{k_{\rm B}}

\newcommand{\vek} [1]{\mbox{\boldmath${#1}$\unboldmath}}

\title[CIB distortions]{Importance of intracluster scattering and relativistic corrections  from tSZ effect with Cosmic Infrared Background}

\begin{document}

\author[S.~K.~Acharya and J.~Chluba]{
Sandeep Kumar Acharya$^1$\thanks{E-mail:sandeep.acharya@manchester.ac.uk}
and 
Jens Chluba$^1$\thanks{E-mail:jens.chluba@manchester.ac.uk}
\\
$^1$Jodrell Bank Centre for Astrophysics, School of Physics and Astronomy, The University of Manchester, Manchester M13 9PL, U.K.}

\date{\vspace{-4mm}
Accepted XXX. Received YYY; in original form ZZZ}

\maketitle

\begin{abstract}
The Sunyaev-Zeldovich effect towards clusters of galaxies has become a standard probe of cosmology. It is caused by the scattering of photons from the cosmic microwave background (CMB) by the hot cluster electron gas. In a similar manner, other photon backgrounds can be scattered when passing through the cluster medium. This problem has been recently considered for the radio and the cosmic infrared background.
Here we revisit the discussion of the cosmic infrared background (CIB) including several additional effects that were omitted before.
We discuss the {\it intracluster} scattering of the CIB and the role of {\it relativistic} temperature corrections to the individual cluster and all-sky averaged signals. We show that the all-sky CIB distortion introduced by the scattering of the photon field was underestimated by a factor of $\simeq 1.5$ due to neglecting the intracluster scattering contribution. The CIB photons can scatter with the thermal electrons of both the parent halo or another halo, meaning that there are two ways to gain energy. Therefore, energy is essentially transferred twice from the thermal electrons to the CIB.
We carefully clarify the origin of various effects in the calculation of the average CIB and also scattered signals.
The single-cluster CIB scattering signal also exhibits a clear redshift dependence, which can be used in cosmological analyses, as we describe both analytically and  numerically.
This may open a new way for cosmological studies with future CMB experiments.
\end{abstract}

\begin{keywords}
Cosmology - Cosmic Microwave Background; Cosmology - Theory 
\vspace{-4mm}
\end{keywords}

\section{Introduction}
In recent times, the Sunyaev-Zeldovich (SZ) effect has proven to be extremely important tool for cosmology and astrophysics in the context of Cosmic Microwave Background (CMB) radiation. Energetic electrons residing inside galaxy clusters boost the CMB photons , giving rise to the thermal SZ (tSZ)  effect \citep{Zeldovich1969}. Similarly, moving electrons can impart energy to CMB photons via doppler shift which is referred to as kinetic SZ (kSZ) effect \citep{Sunyaev1980}. Electrons and photons exchange energy via tSZ effect which imprints a characteristic distortion on the CMB spectrum while Doppler shifts via kSZ do not change the spectral shape of CMB which is a blackbody. The temperature distortion on CMB due to tSZ effect is $\frac{\Delta T}{T}\big|_{\rm tSZ}\simeq -2\frac{k_{
\rm B} T_{
\rm e}}{m_{
\rm e} c^2}\tau$ at low frequencies, while for the kSZ effect one has $\frac{\Delta T}{T}\big|_{\rm kSZ}\simeq\frac{\varv}{c}\tau$, where $\tau$ is the optical depth for photon passing through the electrons inside the structures and $\frac{\varv}{c}$ is the line of sight velocity component. For galaxy clusters, $T_{
\rm e}\simeq 10^7-10^8$ K while $\varv/c\simeq 10^{-3}$. This makes tSZ the dominant effect by an order of magnitude compared to kSZ. For a review on SZ effects, the readers are referred to \citet{Birkinshaw1999} and \citet{SZreview2019}. 

The tSZ effect due to its unique spectral shape has become a routine probe to detect biggest structures in Universe \citep{Planck2016ymap,Bleem2015,Hilton2021}. These analysis assume that Compton scattering between the hot electrons and the CMB photons is non-relativistic, in which case, one can solve for the distortion shape analytically \citep{Zeldovich1969}, yielding the so-called $y$-distortion. The electrons boost the CMB photons from low to high frequency, therefore, there is a deficit of photons at low frequencies with minima at 150 GHz (which is the frequency at which CMB spectrum peaks) and a surplus of photons at higher frequencies with a null point at $\simeq$ 217 GHz. 

Since the electrons inside the galaxy clusters have temperatures of few keV, several works \citep{Challinor1998,Sazonov1998,Itoh98,ChlubaSZpack}  have pointed out the importance of relativistic corrections of Compton scattering, which shifts the null point to higher frequency as more energetic electrons can boost the CMB photons to higher energy. Extracting the relativistic corrections is important next step to accurately quantifying the energy content of electrons inside the galaxy clusters \citep[e.g.,][]{Erler2017, RC2020}, which can have important cosmological implications \citep{RBRC2019}.  

Almost all works in literature deal with the tSZ effect on CMB photons only. However, the same physics should apply for any cosmological radiation background passing through clusters. The kinematics of Compton scattering and, therefore, the spectral distortion shape depends on the spectrum of the incoming radiation. Recently, the scattering of photon from the radio background \citep{HC2021,LCH2022} and the infrared background \citep{SHB2022} were considered. 

Qualitatively, the calculations have multiple similarities with the tSZ from the CMB, but there are important differences as well. While CMB is truly an isotropic background radiation, the same can not be said about radio and cosmic infrared backgrounds. While there are several works discussing the possibility of radio backgrounds and their relation to the radio excess within the frequency band of 0.1$-$10 GHz \citep{Fixsen2011excess,DT2018} and 21cm absorption feature at $z\simeq 20$ \citep{Edges2018}, concrete evidence for the excess background is yet to be established. In this case, the radio SZ effect in combination with the tSZ effect may allow us to extract valuable new information about the properties of this background \citep{HC2021,LCH2022}.

In contrast, \cite{Planck_cib,Melin2018} has shown the existence of the cosmic infrared background (CIB) with theoretical works  \citep{SZKO2012,MM2021} using the halo model \citep{S2000} description to explain the radiation that we see today. Within this model, the radiation originates from galaxies residing inside the dark matter halos which only form at $z\lesssim 6$. As more structures begin to form, the radiation from individual halos starts to build up and become homogeneous. Therefore, the CIB is not really a (primordial) background radiation but evolves over redshifts. 
This results in important difference between tSZ from CMB and CIB. It is well known that the $y$-distortion shape for a galaxy cluster with redshift-independent electron distribution is unique and independent of the redshift at which the CMB distortion was created. This is because redshifting of the SZ spectrum does not affects its shape. However, the same cannot be said about CIB, as the radiation field is itself evolving as structures form. Therefore, one has to compute the tSZ signal from CIB scattering (henceforth referred to as comptonized-CIB or cCIB) from individual objects on a case-by-case basis. This complicates the calculations but also opens an interesting avenue to do tomography using the cCIB effect.    

The authors in \cite{SHB2022} have computed the corresponding CIB distortion signature using the halo model description with evolving CIB. They assumed that halos with mass $M>10^{10}M_{\odot}$ can host radiation emitting galaxies which are already in place at $z\approx 6$. The radiation at such early epochs escapes the parent halo and has the highest probability to scatter with electrons inside the biggest halos which only form around $z\lesssim 1$.  In this work, we show the importance of "intracluster" scattering of infrared radiation, i.e., the scattering of photons within their parent halo. The contribution to the CIB distortion from intracluster scattering is at least as important as "intercluster" contribution with different spectral signature. We emphasize that our calculations are very general and the physics should apply for other evolving radiation backgrounds such as the radio background. We also consider the relativistic temperature corrections to the distortion signal, showing that it can become important for individual clusters.

The paper is structured as follows. In Sec. \ref{sec:halo_model}, we describe our halo model calculations to compute the CIB background. Additionally, we provide analytic approximations to compute the CIB background. We compute the CIB spectral distortion we are expected to see from individual objects as well as average sky signal from all sources in Sec . \ref{sec:SD_shape}. We compute the spectral distortion shapes using the exact Compton scattering kernel and provide fits to null points as a function of CIB spectrum parameters. In Sec. \ref{sec:cluster_scattering}, we compute the intracluster contribution to sky average CIB taking into account the distribution of galaxies inside the dark matter halos. We finish with some discussions and plan for future work in Sec. \ref{sec:conclusions}.

\section{Infrared background with halo model}
\label{sec:halo_model}
We use the halo model \citep{S2000} description of \cite{SZKO2012,MM2021}, which was used to obtain the CIB radiation in \cite{SHB2022}. We assume that the emission of photons from galaxies inside the dark matter halos gives rise to the CIB. The number of galaxies inside a dark matter halo is a function of mass of dark matter halo itself. The emitted photons from galaxies then travel through the expanding universe and show up as the CIB radiation today. 

A detailed derivation of the required radiative transfer solutions given in Appendix~\ref{app:detailed_derive}. The final solution for the ambient CIB at a redshift $z$ and frequency $\nu$ is
\begin{subequations}
\label{eq:sol_ICIB}
\begin{align}
I^{\rm CIB}_\nu(z)&=\frac{1}{a^3}\int^{z_{\rm max}}_z \,\frac{c\,{\rm d} z'}{H(z')}
\,\frac{a'\bar{L}^{\rm h}(z')}{4\pi}
\,\Theta_{\nu'}(z'),
\\
a\,\bar{L}^{\rm h}(z)&=\int {\rm d}M\,\frac{{\rm d}N(M,z)}{{\rm d}M}\,a\,L^{\rm h}(M,z)
\end{align}
\end{subequations}
with $\nu'=\nu\,a/a'$ and where we factored the frequency-dependence, $\Theta_{\nu'}(z')$, of the halo luminosity, $L^{\rm h}_\nu(M,z)$, as we describe below and  ${\rm d}N/{\rm d}M$ is the halo mass function. For convenience, we also introduced the mass-function-averaged luminosity $\bar{L}^{\rm h}_\nu(z)$. We use $z_{\rm max}=6$ as the maximal emission redshift within the halo model, finding that beyond that little is added. 
Comparing to \citet{SHB2022}, here we give the solution for the physical intensity at every redshift, which gives rise to an extra factor of $1/a^3=(1+z)^3$. 
To give the ambient CIB at the location of the cluster, we now compute $\bar{L}^{\rm h}_\nu(z)$ using the halo model.

\subsection{Halo model for CIB radiation}
In the halo model, the galaxy luminosity of a halo with mass $M$ at redshift $z$ is given by,
     $L_{\nu}^{\rm gal}(M,z)=L_0\,\Phi(z)\,\Sigma(M)\,\Theta_\nu(z)$,
with normalization $L_0=6.4\times 10^{-8}$ Jy Mpc$^2=6.09\times 10^{11}$ W\,Hz$^{-1}$, $\Phi(z)=(1+z)^{3.6}$, and
\begin{align}
    \Sigma(M)=\frac{M/M_\odot}{\sqrt{2\pi\sigma^2}}{\rm e}^{-\frac{\log^2_{10}(M/M_{\rm eff})}{2\sigma^2}},
\end{align}
with $M_{\rm eff}=10^{12.6}M_{\odot}$ and $\sigma^2=0.5$. 
The spectral energy distribution (SED), $\Theta_\nu(z)$, of the emitting galaxies as a function of rest frame frequency $\nu$ and $z$ is given by a modified blackbody spectrum
\begin{equation}
\label{eq:define_theta}
 \Theta_\nu(z)=
 \begin{cases}
\left(\frac{\nu}{\nu_0}\right)^{\beta}\frac{B_{\nu}[T_{\rm d}(z)]}{B_{\nu_0}[T_{\rm d}(z)]}
&\text{for $\nu\leq\nu_0$}
\\[2mm]
\left(\frac{\nu}{\nu_0}\right)^{-\gamma}
&\text{for $\nu>\nu_0$}.
\end{cases}
\end{equation}
Here, $B_{\nu}(T_{\rm d})$ is the blackbody SED at temperature $T_{\rm d}(z)=T_0(1+z)^{\alpha}$ with $\alpha=0.36$ and $T_0=24.4$ K. The pivot frequency is given by
\begin{equation}
\label{eq:pivot}
    \nu_0(z)=\frac{k_{\rm B} T_{\rm d}(z)}{h}\bigg[3+\beta+\gamma+W_0(\lambda)\bigg]
\end{equation}
with $\lambda=-(3+\beta+\gamma)\,{\rm e}^{-(3+\beta+\gamma)}$, where $W_0$ is the Lambert function with $\beta=1.75$ and $\gamma=1.7$. 

The total galaxy luminosity of a halo has two main contributions. The first is from the central galaxy,
\begin{equation}
L_{\nu}^{\rm c}(M,z)=N^{\rm c}(M,z)\,L_{\nu}^{\rm gal}(M,z),
\end{equation}
where $N^{\rm c}$ is the number of central galaxies inside the dark matter halo. In our model it is 1 for $M> 10^{10}M_{\odot}$ and zero otherwise. 

The second contribution to the halo luminosity is from the satellite galaxies, which is determined by
\begin{equation}
    L_{\nu}^{\rm s}(M,z)=\int_{M_{\rm min}}^M {\rm d}M_{\rm s}\,\frac{{\rm d}N_{\rm s}(M,M_{\rm s})}{{\rm d}M_{\rm s}}\,L_{\nu}^{\rm gal}(M_{\rm s},z),
\end{equation}
where ${\rm d}N_{\rm s}/{\rm d}M_{\rm s}$ is the sub-halo mass function, i.e., the number of sub-halos as a function of host halo mass. We use the expressions of \citet{TW2010}
with lower mass limit $M_{\rm min}=10^{10}M_{\odot}$. 

The total halo luminosity is then $L^{\rm h}_{\nu}(M,z)=L_{\nu}^{\rm c}(M,z)+L_{\nu}^{\rm s}(M,z)$. It is useful to introduce $L^{\rm h}(M,z)=\frac{L^{\rm h}_{\nu}(M,z)}{\Theta_\nu(z)}$ as
\begin{align}
    &L^{\rm h}(M,z)
    =
    L_0\,\Phi(z)\,\Bigg[N^{\rm c}(M,z)\,\Sigma(M)
    +\int_{M_{\rm min}}^M \!{\rm d}M_{\rm s}\,\frac{{\rm d}N_{\rm s}(M,M_{\rm s})}{{\rm d}M_{\rm s}}\,\Sigma(M_{\rm s})\Bigg]
    \nonumber
\end{align}
to cleanly separate frequency- and mass-dependent terms. 
We then have the mass-function-averaged luminosity
\begin{equation}
    \bar{L}^{\rm h}(z)
    = \int_{M_{\rm min}}^{M_{\rm max}} {\rm d}M\,\frac{{\rm d}N(M,z)}{{\rm d}M}\,L^{\rm h}(M,z).
    \label{eq:averaged_source_term_LM}
\end{equation}
For the halo mass function, we use the fits provided by 
\citet{TKKAWYGH2008}. The halo mass functions are defined as overdensity w.r.t either the critical energy density or mass energy density. To convert the mass definitions we use the concentration mass relations of \citet{DSKD2008} and \citet{HK2003}. We choose $M_{\rm min}=10^{10}M_{\odot}$ and $M_{\rm max}=10^{16}M_{\odot}$ in our computations.
%

\begin{figure}
\centering 
\includegraphics[width=\columnwidth]{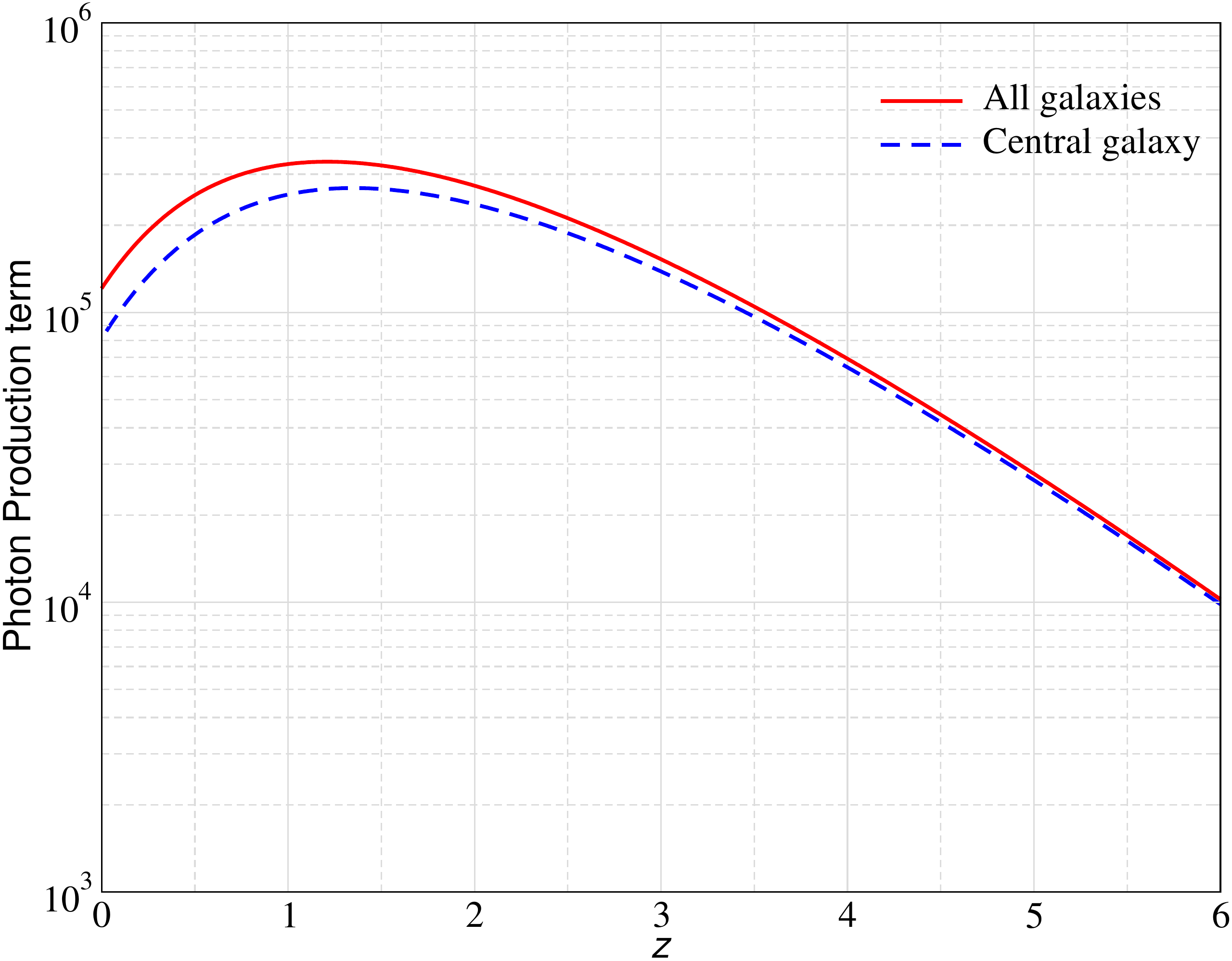}
\\
\caption{Illustration of $\mathcal{S}(z)=c\,a\,\bar{L}^{\rm h}(z)/[4\pi\, H(z)]$ in Jy/Sr, which describes the CIB photon source term as a function of redshift (see Appendix~\ref{app:fit_Lz} for fit).
The solid line includes both sub-halo and central galaxy contributions, while the dashed line only accounts for the central galaxy term.
Most photons originate from $z\simeq 1$, while at $z\gtrsim 6$ very few photons are generated.}
\label{fig:Lbar}
\end{figure}

\begin{figure}
\centering 
\includegraphics[width=0.96\columnwidth]{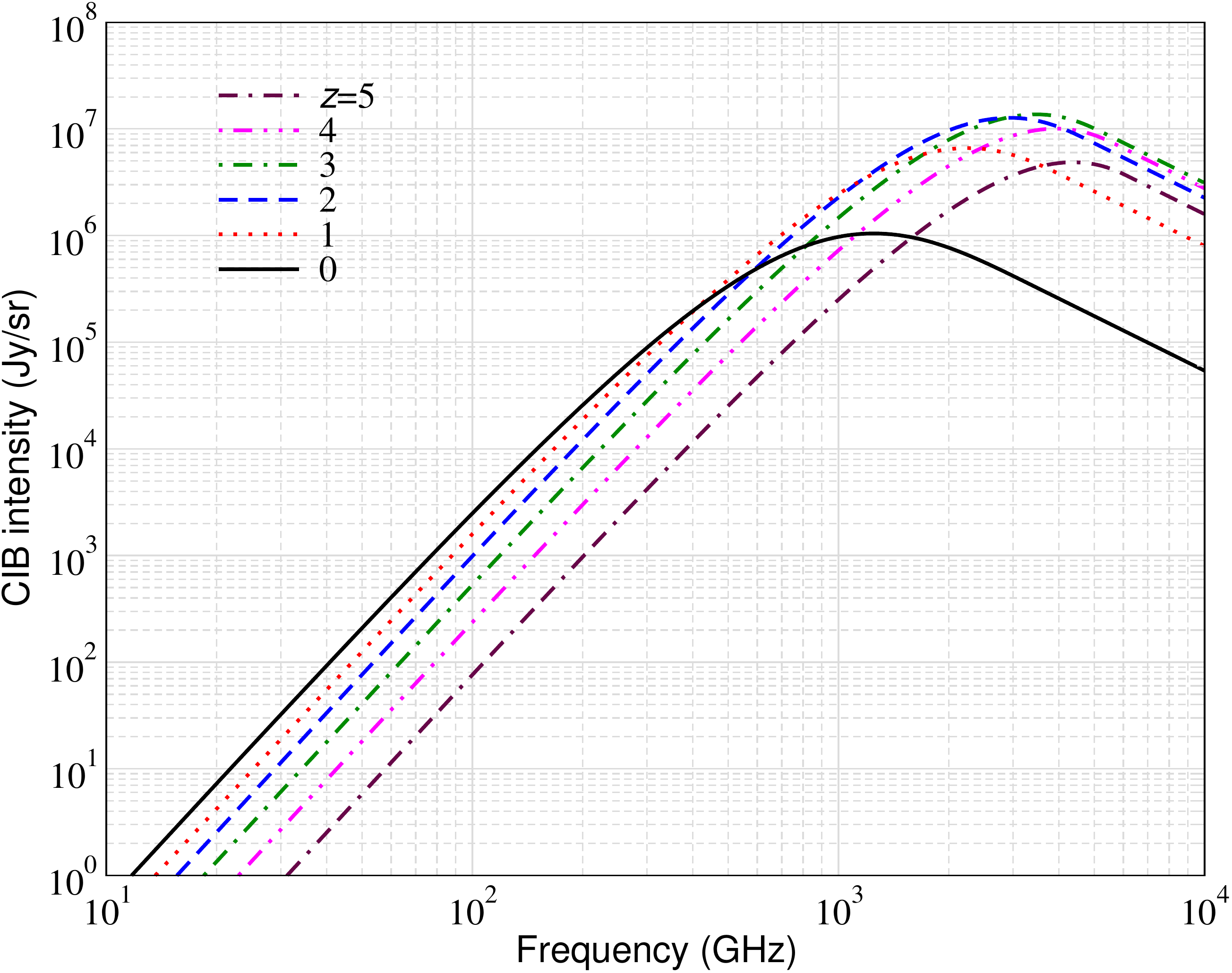}
\\[2mm]
\includegraphics[width=0.96\columnwidth]{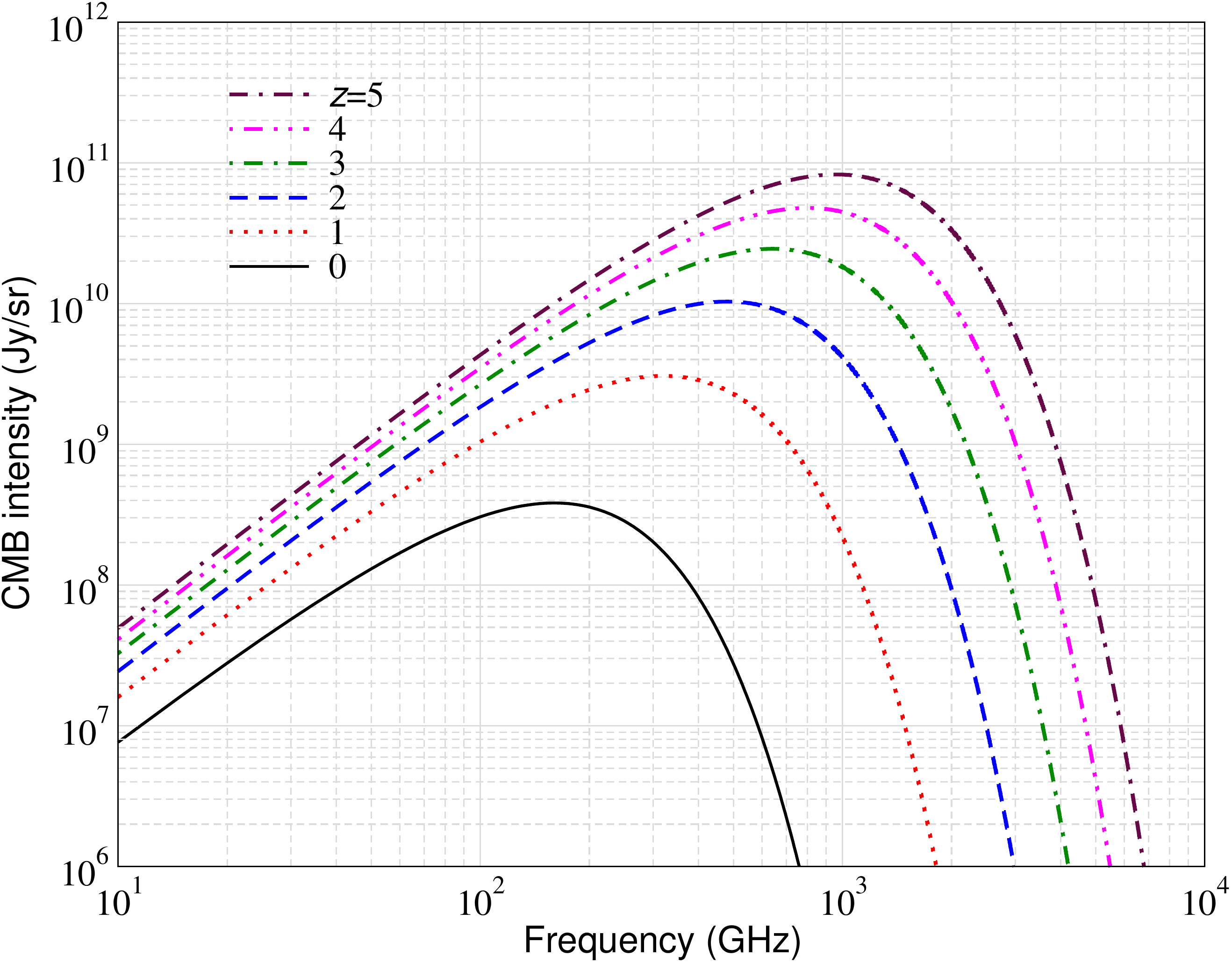}
\\
\caption{Physical intensity of the ambient CIB (upper panel) as a function of redshift. The expansion of the Universe and evolution of the CIB temperature change the position and amplitude of the signal. For comparison we also show the same result for the CMB (lower panel), which changes more significantly with redshift due to the more strong evolution of the CMB temperature [$T\simeq (1+z)$ instead of $T\simeq (1+z)^{0.36}$] and drop in the number of CIB emitters towards high $z$.}
\label{fig:CIB_spectrum}
\end{figure}

\subsection{Average CIB as a function of redshift}
We now have all the ingredients to compute the average CIB spectrum. In Fig.~\ref{fig:Lbar}, we illustrate the CIB photon source term defined as $\mathcal{S}(z)=c\,a\,\bar{L}^{\rm h}(z)/[4\pi\, H(z)]$, which directly appears in the integral of Eq.~\eqref{eq:sol_ICIB}. The precise redshift-scaling depends on the halo-model parameters and the cosmological model, however, we do not vary them here.
Most of the CIB photons are produced at $z\simeq 1$, where star-formation is efficient. The fall-off towards higher redshifts means that the ambient CIB becomes dimmer. We note that at all redshifts the central galaxy contribution dominates.

In Fig. \ref{fig:CIB_spectrum}, we show the evolution of physical intensity of the CIB and the CMB as a function of frequency for several redshifts. This was computed numerically with Eq.~\eqref{eq:sol_ICIB} and can be thought of as the ambient CIB light that an observer inside a cluster at the corresponding redshift would receive.
We use $z_{\rm max}=6$ in our computations. 
Towards higher redshifts the effective temperature of the CIB increases as $T\propto (1+z)^{0.36}$, resulting in a shift of the peak position towards higher frequencies. This shift is smaller than for the CMB (see lower panel in Fig. \ref{fig:CIB_spectrum}), for which the peak position scales as $\propto (1+z)$.

At low frequency, the CIB and CMB intensity behave in opposite ways. This is due to the fact that the CIB only slowly builds up as a function of time, with most of the CIB being created at $z\simeq 1$. The energy density of CMB keeps on decreasing with decreasing redshift and the intensity is proportional to $\simeq (1+z)$.
In addition, we can see that at low frequencies the brightness of the ambient CIB remains nearly constant and then drops off at $z\gtrsim 1$ while that of CMB increases as $\simeq (1+z)$, as already mentioned. Knowing that scattering of the ambient CMB as shown in Fig.~\ref{fig:CIB_spectrum} leads to a signal that, after redshifting to $z=0$, is independent of the scattering redshift, we can anticipate that the scattered CIB signal for each cluster will depend on the scattering redshift.

\begin{figure}
\centering 
\includegraphics[width=\columnwidth]{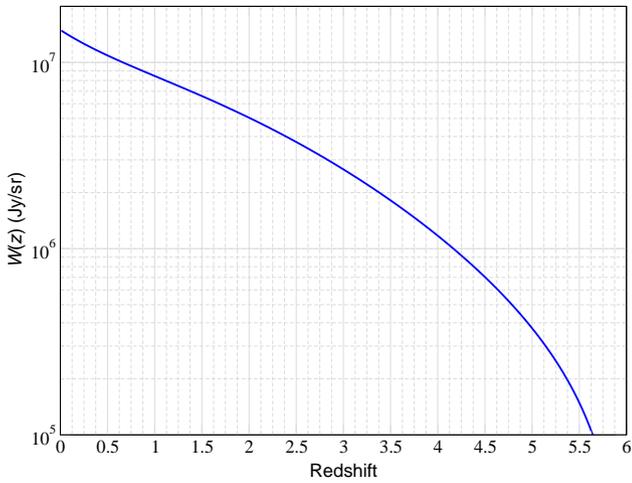}
\\
\caption{Redshift dependence of the integral $W(z)=(1+z)^{1.65}\mathcal{J}(z, 1-\alpha)$, relevant to Eq.~\eqref{eq:sol_ICIB_low} for the low-frequency approximation of the CIB.}
\label{fig:special_int}
\end{figure}

\subsubsection{Analytic approximations for the average CIB}
\label{sec:ana_CIB}
To obtain a more quantitative understanding of the average CIB spectrum at different redshifts, we can use a moment-expansion approach \citep{Chluba2017foregrounds} to approximate its evolution. For this we rewrite the fundamental SED in Eq.~\eqref{eq:define_theta}. Since according to Eq.~\eqref{eq:pivot}, the dimensionless pivot frequency,
\begin{equation}
\label{eq:x0_def}
 x_0=\frac{h\nu_0}{k_{\rm B}T_{\rm d}}=
 3+\beta+\gamma+W_0(\lambda)\approx 6.440,
\end{equation}
is independent of redshift, it is convenient to write
\begin{equation}
\label{eq:define_theta_mod}
 \Theta_\nu(z)=
 \begin{cases}
\left(\frac{x(z)}{x_0}\right)^{\beta+3}\frac{{\rm e}^{x_0}-1}{{\rm e}^{x(z)}-1}
&\text{for $x(z) \leq x_0$}
\\[2mm]
\left(\frac{x(z)}{x_0}\right)^{-\gamma}
&\text{for $x(z)>x_0$}.
\end{cases}
\end{equation}
with $x(z)=h\nu/k_{\rm B}T_{\rm d}(z)$. In Eq.~\eqref{eq:sol_ICIB}, we have to evaluate the SED at $\nu'=\nu a/a'$ and $z'$, implying $x'(z')=h\nu'/k_{\rm B}T_{\rm d}(z')=x(z)\,(a'/a)^{\alpha-1}$. Since the main SED scales as $\Theta_{\nu'}(z')\propto \left(a'/a\right)^{(\alpha-1)(\beta+3)}$, for the moment treatment it is useful to define the integrals
\begin{equation}
\label{eq:def_X_av}
    \langle X \rangle
    = \int^{z_{\rm max}}_z \frac{c\,{\rm d} z'}{H(z')}
    \,\frac{a'\bar{L}^{\rm h}(z')}{4\pi}
    \,\left(\frac{a'}{a}\right)^{(\alpha-1)(\beta+3)}\,X(z').
\end{equation}
of a quantity $X(z')$. These averages can be quickly computed using the fit in Appendix~\ref{app:fit_Lz}.
Since we will encounter multiple cases of the form $\langle (a'/a)^p \rangle$, we also introduce the function
$\mathcal{J}(z, p)=\langle (a'/a)^p \rangle$.
At low-frequencies (i.e., $x'(z')\ll 1$), with Eq.~\eqref{eq:sol_ICIB} we then find
\begin{equation}
\label{eq:sol_ICIB_low}
    I^{\rm CIB}_\nu(z)
    \approx \frac{1}{a^3}\,
    \left(\frac{x(z)}{x_0}\right)^{\beta+2}
    \left< \frac{x(z)}{x'(z')} \right>
    =\frac{1}{a^3}\,
    \left(\frac{x(z)}{x_0}\right)^{\beta+2}
    \,\mathcal{J}(z, 1-\alpha).
\end{equation}
Although $(1+z)^3 x(z)^{\beta+2}\propto (1+z)^{3-\alpha(\beta+2)}\simeq (1+z)^{1.65}$ increases with redshift, the remaining integral in Eq.~\eqref{eq:sol_ICIB_low} weakens the rise, leading to an overall drop of the intensity with redshift. 
In Fig~\ref{fig:special_int}, we illustrate the redshift-dependence of this integral.
From this, we expect a drop of the CIB by a factor of $\simeq 1.8$ between $z=0$ and $z=1$ and a drop by a factor of $\simeq 40$ between $z=0$ and $z=5$. This is consistent with what is shown in Fig.~\ref{fig:CIB_spectrum} at $\nu\lesssim 100\,{\rm GHz}$.

At very high frequency (i.e., $x'(z')> x_0$ at all $z'$) it is also quite easy to provide a simple approximation. Using the notation from above we find
\begin{equation}
\label{eq:sol_ICIB_high}
    I^{\rm CIB}_\nu(z)
    \approx
    \frac{1}{a^3}\,
    \left(\frac{x(z)}{x_0}\right)^{-\gamma}
    \mathcal{J}(z,[1-\alpha][\gamma+3+\beta]).
\end{equation}
to work extremely well.
However, what do we do about the intermediate regime? In particular, it turns out that this is where the CIB maximum is located and estimates neglecting the high-frequency CIB SED contributions do not give accurate results. To tackle the problem, we realize that for a given frequency $\nu$ and observing redshift $z$, there is a critical redshift $z_{\rm c}$ that allows us to write
\begin{align}
\label{eq:sol_ICIB_split}
    I^{\rm CIB}_\nu(z)&=\frac{1}{a^3}\Bigg\{
    \int^{\min(z_{\rm c},z_{\rm max})}_z  \,\frac{c\,{\rm d} z'}{H(z')}
    \,\frac{a'\bar{L}^{\rm h}(z')}{4\pi}
    \,\left(\frac{x'(z')}{x_0}\right)^{\beta+3}\frac{{\rm e}^{x_0}-1}{{\rm e}^{x'(z')}-1}
    \nonumber\\
    &\quad\qquad+
    \int_{\max(z,z_{\rm c})}^{z_{\rm max}}  \,\frac{c\,{\rm d} z'}{H(z')}
    \,\frac{a'\bar{L}^{\rm h}(z')}{4\pi}
    \,\left(\frac{x'(z')}{x_0}\right)^{-\gamma}
    \Bigg\}.
\end{align}
The critical redshift is determined by the condition $x'(z')=x_0$:
\begin{align}
\label{eq:zcrit}
1+z_{\rm c}=\left(\frac{x_0}{x(z)}\right)^{\frac{1}{1-\alpha}}(1+z)=\left(\frac{3274.2\,{\rm GHz}}{\nu a} \right)^{\frac{1}{1-\alpha}}.
\end{align}
Whenever $z_{\rm c}>z_{\rm max}$, only the first integral contributes. Demanding $z_{\rm c}=z_{\rm max}$ then determines the critical frequency
\begin{align}
\nu_{\rm c}(z)=3274.2\,{\rm GHz}\frac{(1+z)}{(1+z_{\rm max})^{1-\alpha}}\approx 942.4\,{\rm GHz} \,(1+z)
\end{align}
below which the second integral can be neglected. We can also demand $z_{\rm c}=z$, from which we find 
$\nu_{\rm h}(z)=3274.2\,{\rm GHz} \,(1+z)^{\alpha}$
above which the high-frequency solution, Eq.~\eqref{eq:sol_ICIB_high}, is applicable.

While $\nu\leq \nu_{\rm c}(z)$, we can focus on the first integral in Eq.~\eqref{eq:sol_ICIB_split}. To simplify matters, we first write $x'(z')=x(z)\,\xi$ with $\xi=(a'/a)^{\alpha-1}$. We next expand the Planckian $1/({\rm e}^{x'(z')}-1)$ around some average value $\xi^*$, which we still need to determine:
\begin{equation}
\nonumber
\frac{1}{{\rm e}^{x'(z')}-1}
 \approx \frac{1}{{\rm e}^{x^*}-1}\left\{1-G(x^*)
 \left[\frac{\xi}{\xi^*}-1\right]
 +\frac{1}{2}\,Y^*(x^*) \left[\frac{\xi}{\xi^*}-1\right]^2\right\}.
\end{equation}
Here, $x^*=x\,\xi^*$, and where we introduced the function $G(x)=\frac{x {\rm e}^{x}}{({\rm e}^{x}-1)}$ and $Y^*(x)=G(x)\,x\coth\left(\frac{x}{2}\right)$. Introducing the normalization $\bar{A}=\left<1\right>$, we then define the redshift moments
\begin{align}
\label{eq:moments_CIB}
M_k &= \bar{A}^{-1}\left<\left[\frac{\xi}{\xi^*}-1\right]^k\right>.
\end{align}
These can be quickly computed using the approximation Eq.~\eqref{eq:Sfit} with Eq.~\eqref{eq:def_X_av}.
By demanding that the first moment vanishes we can then determine the pivot value as
\begin{align}
\label{eq:xi_CIB}
\xi^*(z) &= \mathcal{J}(z, \alpha-1)/\bar{A}\equiv
\mathcal{J}(z, \alpha-1)/\mathcal{J}(z, 0).
\end{align}
A fit for this function is given in Appendix~\ref{app:fitxi}. At $\nu\leq \nu_{\rm c}(z)$, we then have the final approximation
\begin{align}
\label{eq:sol_ICIB_max}
    I^{\rm CIB}_\nu(z)
    &\approx 
    \frac{\bar{A}}{a^3}\,\left(\frac{x}{x_0}\right)^{\beta+3}\frac{{\rm e}^{x_0}-1}{{\rm e}^{x^*}-1}
    \,\Bigg[1
    +\frac{1}{2}\,Y^*(x^*)\,M_2(z).
    \Bigg].
\end{align}
Indeed we find that the second moment correction remains noticeable close to $\nu_{\rm c}(z)$ and at low $z$. We also confirmed that adding higher order moment terms does not improve the match much.

If we now enter the regime $\nu_{\rm c}(z)<\nu\leq \nu_{\rm h}(z)$, we can equally use Eq.~\eqref{eq:sol_ICIB_max}; however, in the integral Eq.~\eqref{eq:def_X_av}, we then have to use the upper boundary $z_{\rm c}$ instead of $z_{\rm max}$. In addition we have to add the high frequency solution, Eq.~\eqref{eq:sol_ICIB_high} but with the modified lower boundary (i.e., $z\rightarrow z_{\rm c}$) in the corresponding integral. Since in this regime the integrals directly depend on the chosen frequency, one can no longer pre-compute the solutions. Therefore, the approximation becomes moot and the full integral expression should be used.


\begin{figure}
\centering 
\includegraphics[width=\columnwidth]{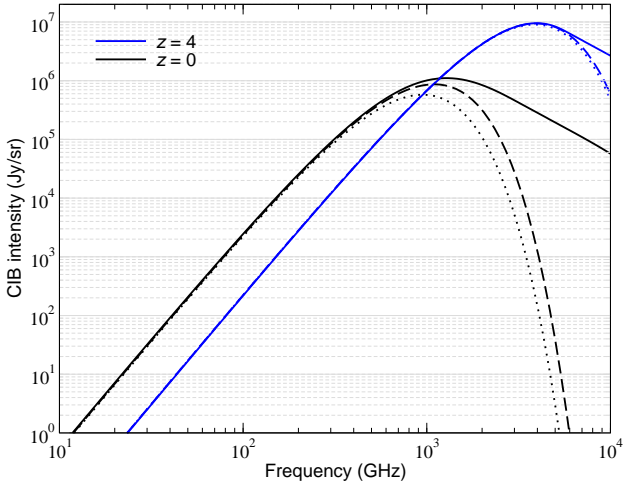}
\\
\caption{CIB spectrum at $z=0$ and $z=4$. The solid lines show the exact result, while the dashed and dotted lines respectively show the approximation in Eq.~\eqref{eq:sol_ICIB_max} with and without the second moment term included.}
\label{fig:CIB_spectrum_approx}
\end{figure}
In Fig.~\ref{fig:CIB_spectrum_approx} we illustrate the performance of the approximation given by Eq.~\eqref{eq:sol_ICIB_max}. We chose $z=0$ and $z=4$ as examples. For these redshifts we have $(\bar{A},\xi^*, M_2)\approx (3.632\times 10^{7}\,{\rm Jy/sr}, 2.544, 0.0389)$ and 
$(\bar{A},\xi^*, M_2)\approx (9.074\times 10^{4}, 1.105, 0.00373)$, respectively.
As expected, the approximation fails in the Wien tail of the CIB, while below the respective critical frequencies, $\nu_{\rm c}\approx 942.4\,{\rm GHz}$ and $\nu_{\rm c}\approx 4712\,{\rm GHz}$, it works very well. At low $z$, we also find the second moment term to be noticeable, while it becomes small early on. We also note that setting $\xi^*=1$ would not reproduce the CIB SED for $z=0$.

\section{Spectral distortion shapes from scattering of CIB photons with thermal electrons}
\label{sec:SD_shape}
The free-streaming CIB photons encounter hot electrons inside the dark matter halos while travelling towards us. As a result, the electrons impart energy to the photons via inverse Compton scattering, usually boosting them from lower to higher frequency. This results in distortion in spectral shape of photon radiation field, which was recently considered by \citet{SHB2022} using the Kompaneets equation \citep{Kompa56}. 

In this section, we briefly repeat the derivation of \citet{SHB2022} and then extend it by considering the scattering process using the exact Compton scattering kernel \citep{CSpack2019}. This allows us to include relativistic corrections to the scattering process, which changes the CIB scattering distortion for each cluster.

\subsection{Kompaneets treatment for single clusters}
\label{sec:SD_shape_Kompaneets}
The authors of \citet{SHB2022} have calculated the spectral distortion shape by solving Kompaneets equation \citep{Kompa56}, which assumes the scattering between the hot electrons and the photons to be non-relativistic. Since the typical energy of the background photons is low compared to the electron energy ($h\nu\ll k_{\rm B} T_{\rm e}$), one can neglect stimulated scattering and electron recoil effects.
The scattering of the average radiation field by the thermal electrons inside the cluster can then be described as \citep{Zeldovich1969, Sazonov2000}
\begin{align}
\frac{{\rm d}n_\nu}{{\rm d} y_{\rm sc}}\approx\frac{1}{\nu^2}\frac{\partial}{\partial \nu} \nu^4 \frac{\partial}{\partial \nu} n_\nu,
\end{align}
where $n_\nu=I_\nu/[2h\nu^3]$ is the photon occupation number and $y_{\rm sc}$ is the scattering $y$-parameter. We use the optically-thin limit, such that the scattered radiation can be estimated as 
\begin{align}
\Delta I^{\rm CIB}_\nu(z_{\rm sc})\approx y_{\rm sc}\nu\,\frac{\partial}{\partial \nu} \nu^4 \frac{\partial}{\partial \nu} \frac{I^{\rm CIB}_\nu(z_{\rm sc})}{\nu^3}
\end{align}
after passing through the cluster at the scattering redshift $z_{\rm sc}$ (or scale-factor $a_{\rm sc}=1/[1+z_{\rm sc}]$). Using Eq.~\eqref{eq:sol_ICIB} to describe the incoming CIB background, we then simply have
\begin{align}
\label{eq:sol_ICIB_sc}
    \Delta I^{\rm CIB}_\nu(z_{\rm sc})&\approx\frac{y_{\rm sc}}{a_{\rm sc}^3}\!\int^{z_{\rm max}}_{z_{\rm sc}}  \frac{c\,{\rm d} z'}{H(z')}
    \,\frac{a'\bar{L}^{\rm h}(z')}{4\pi}
    \left[\nu\frac{\partial}{\partial \nu} \nu^4 \frac{\partial}{\partial \nu} \frac{\Theta_{\nu'}(z')}{\nu^3}
    \right]
\end{align}
at the scattering redshift.
The scattering operator acts only on $\Theta_{\nu'}(z')$, where $\nu'=\nu\, a_{\rm sc}/a'$. Since we have the identity
\begin{align}
\label{eq:sol_ICIB_sc_op}
   \Delta \Theta_{\rm sc}(\nu', z')
   =
   \nu\frac{\partial}{\partial \nu} \nu^4 \frac{\partial}{\partial \nu} \frac{\Theta_{\nu'}(z')}{\nu^3}
   \equiv\nu'\frac{\partial}{\partial \nu'} {\nu'}^4 \frac{\partial}{\partial \nu'} \frac{\Theta_{\nu'}(z')}{{\nu'}^3},
\end{align}
Using Eq.~\eqref{eq:define_theta}, we find
\begin{equation}
\label{eq:define_theta_sc}
\frac{\Delta \Theta_{\rm sc}(\nu', z')}{\Theta(\nu',z')}=
\begin{cases}
Y(x')-2 \beta \,G(x')+\beta (3+\beta)
&\text{for $\nu'\leq\nu_0$}
\\[2mm]
\gamma (3+\gamma)
&\text{for $\nu'>\nu_0$},
\end{cases}
\end{equation}
with $x'=h\nu'/k_{\rm B} T_{\rm d}(z')$. Here, $Y(x)=G(x)\left[x\coth\left(\frac{x}{2}\right)-4\right]$ is simply the $y$-type distortion caused on the blackbody part of the CIB spectrum and $G(x)=\frac{x\,{\rm e}^{x}}{{\rm e}^{x}-1}$ is the related blackbody temperature shift.

After the scattering occurred, the scattering signal simply evolves according to Eq.~\eqref{eq:evol_transformed} without any new sources towards $z=0$. This includes redshifting the frequency from $z_{\rm sc}$ to $z=0$ by the replacement $\nu\rightarrow \nu /a_{\rm sc}$, and renormalizing the total intensity by a factor of $a_{\rm sc}^3$. At $z=0$, we then have the scattered CIB contribution
\begin{align}
\label{eq:sol_ICIB_sc_z0}
    \Delta I^{\rm CIB}_{\nu, 0}
    &\approx y_{\rm sc}\int^{z_{\rm max}}_{z_{\rm sc}} \!\frac{c\,{\rm d} z'}{H(z')}
    \,\frac{a'\bar{L}(z')}{4\pi}
    \,\Delta \Theta_{\rm sc}(\nu/a', z')
\end{align}
in the direction of the cluster located at $z_{\rm sc}$. This result is in agreement with that given by \citet{SHB2022}. We will use this result for reference; however, for typical clusters seen by {\it Planck} relativistic corrections become important as we discuss next.

\subsection{Compton scattering kernel treatment for single clusters}
\label{sec:SD_shape_Kernel}
The derivations of Sect.~\ref{sec:SD_shape_Kompaneets} assumed that the scattered CIB signal at the redshift $z_{\rm sc}$ can be obtained used the Kompaneets operator. However, the precise spectral distortion shape is determined by the kinematics of the scattering process and therefore not only depends on the SED of the ambient radiation field, but also the energy distribution of the electrons. Since the electrons inside galaxy clusters have energy $\simeq$ few keV, relativistic corrections become important, which change the spectral distortion shapes \citep{Wright1979,Fabbri1981,Rephaeli1995,Challinor1998,Sazonov1998,Itoh98}. One can capture these corrections by performing a Taylor series expansion in $\theta_{\rm e}=k_{\rm B} T_{\rm e}/m_{\rm e}c^2$ \citep{Challinor1998,Sazonov1998,Itoh98}. This can in principle be quickly carried out using {\tt SZpack} \citep{ChlubaSZpack}. For an application of this method to scattering of radio photons by the hot electrons see \citet{LCH2022}.

However, it is well-known that for the scattering of the CMB this approach does not yield accurate results at high electron temperatures or high observing frequencies \citep{Itoh98, ChlubaSZpack}. 
In this work, we therefore compute the spectral distortion shape by solving Boltzmann equation using the exact Compton scattering kernel, $P_{\nu\rightarrow \nu'}$, which can be computed efficiently as a function of the electron temperature using {\tt CSpack} \citep{CSpack2019}. 

Ignoring stimulated scattering effects, the Boltzmann equation can be schematically written as \citep[see][for discussions of numerical methods]{Acharya2021FP},
\begin{align}
\label{eq:kernel_eq}
\frac{{\rm d}I_{\nu_i}}{{\rm d}\tau}\approx
\sum_j\left\{\left(\frac{\nu_i}{\nu_j}\right)^3[P_{\nu_j\rightarrow \nu_i}\Delta \nu] \,I_{\nu_j}-[P_{\nu_i\rightarrow \nu_j}\Delta \nu] \,I_{\nu_i}\right\},
\end{align}
where $P_{\nu_i\rightarrow \nu_j} \Delta\nu$ is the probability of a photon being transferred from frequency bin $i$ to $j$. The inverse scattering event is given by the detailed balanced relation
$P_{\nu_i\rightarrow \nu_j}=(\nu_j/\nu_i)^2{\rm e}^{h(\nu_i-\nu_j)/k_{\rm B} T_{\rm e}}P_{\nu_j\rightarrow \nu_i}$.
The photon number is identically conserved, i.e. $\sum_j P_{\nu_i\rightarrow \nu_j}\Delta \nu=1$. The optical depth is given by $\tau=\int \sigma_{\rm T} N_{\rm e} {\rm d}l$, where $\sigma_{\rm T}$ is Thomson scattering cross-section, $N_{\rm e}$ is the electron number density and ${\rm d}l$ is the differential line segment through the cluster along the line of sight. We assume that the electrons follow a relativistic Maxwell-Boltzmann distribution. 

By inserting the ambient CIB spectrum, Eq.~\eqref{eq:sol_ICIB}, evaluated at the clusters redshift $z_{\rm sc}$ into Eq.~\eqref{eq:kernel_eq}, we can directly evaluate the scattered CIB radiation at $z_{\rm sc}$. Like in the previous section, we then simply replace $\nu\rightarrow \nu / a_{\rm sc}$ and multiply the solution by $a_{\rm sc}^3$ to obtain the final scattering signal. With this two-step procedure, we can obtain the scattered CIB contribution of single clusters residing at various redshifts. Alternatively, we could compute the scattered signal of the fundamental SED, $\Theta(\nu', z')$, before carrying out the integrals over the source redshifts and masses. However, computationally, this is more expensive, since the scattered radiation has to be evaluated for every frequency, redshift and electron temperature.


\begin{figure}
\centering 
\includegraphics[width=0.95\columnwidth]{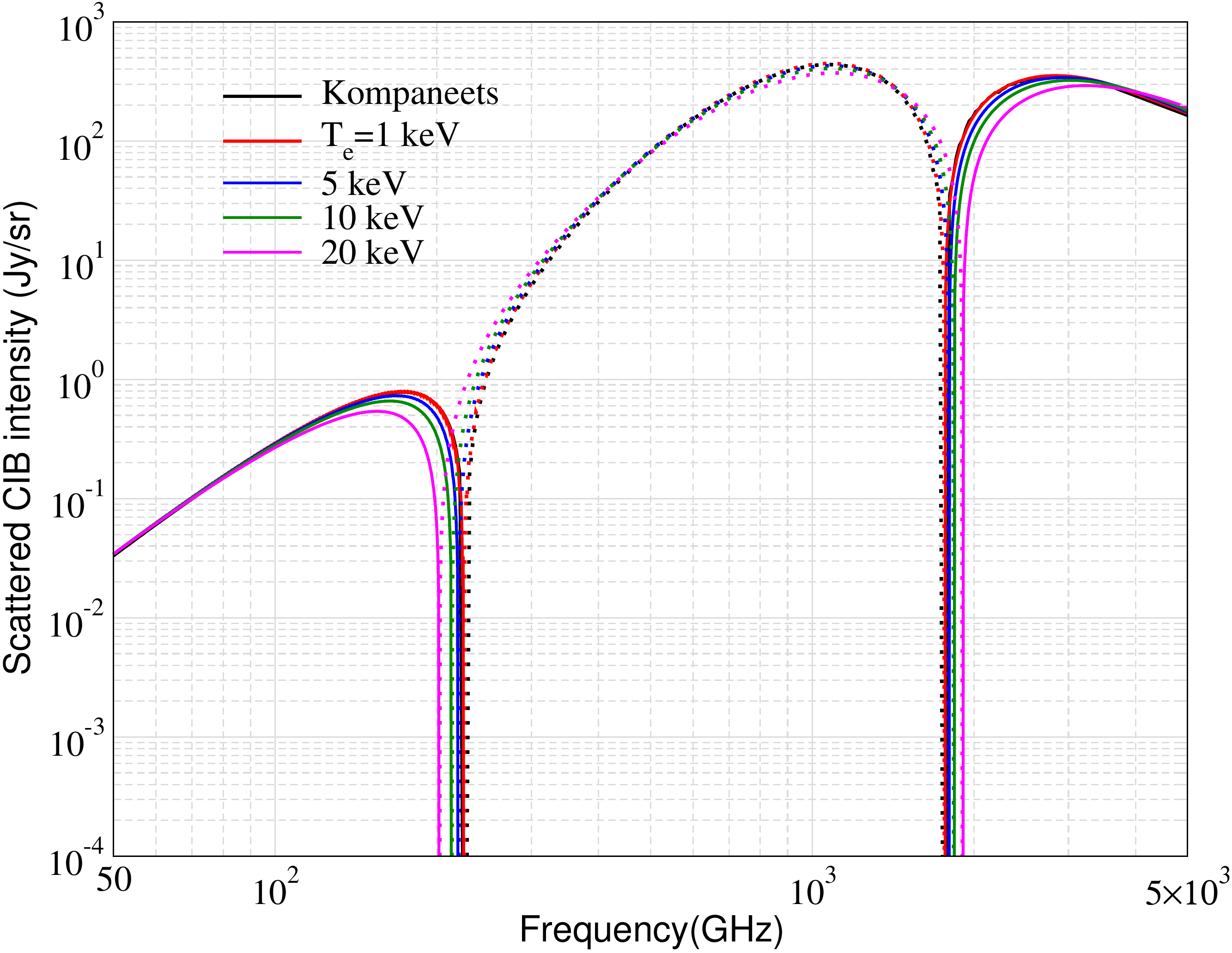}
\\[2mm]
\includegraphics[width=0.95\columnwidth]{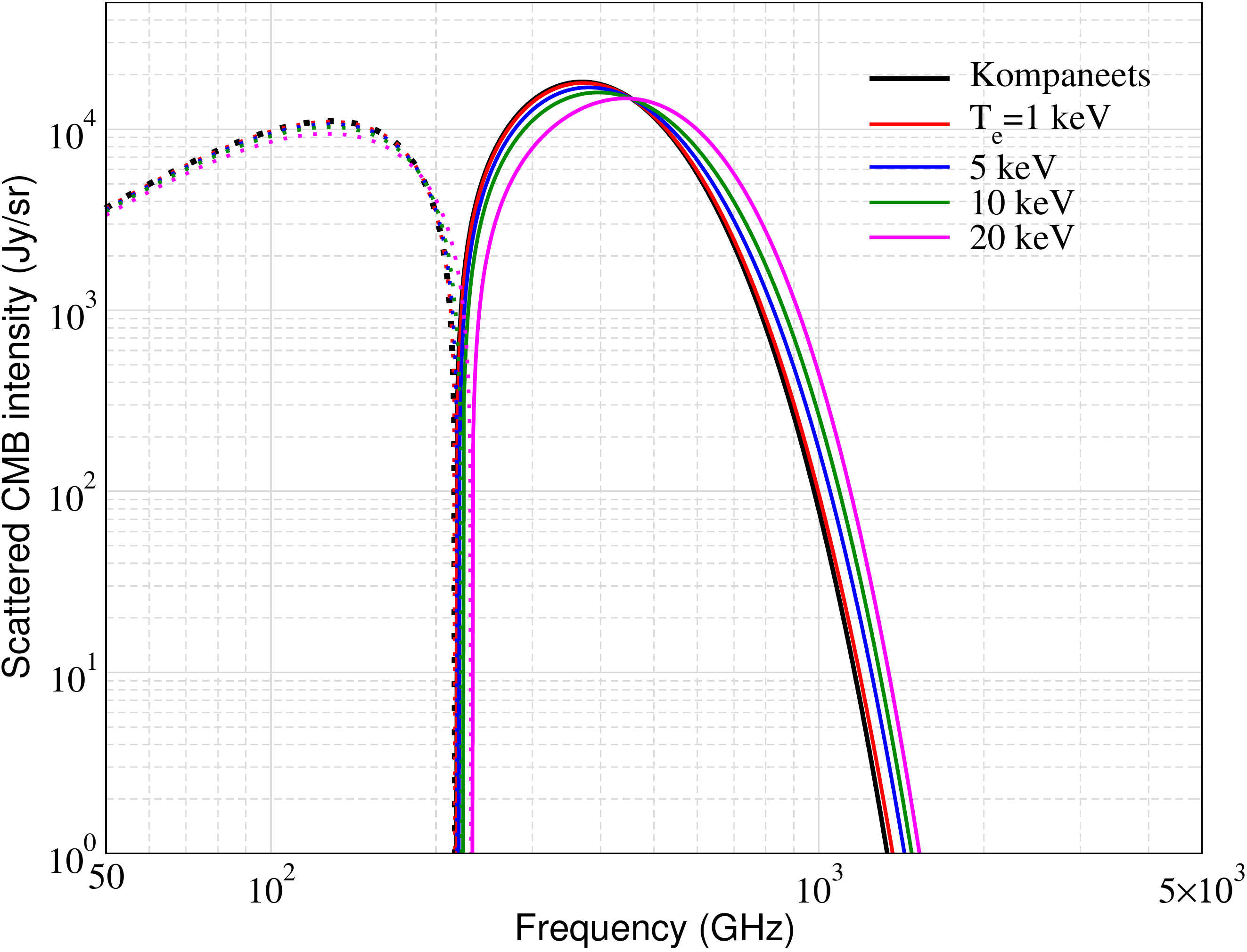}
\\
\caption{CIB (upper panel) and CMB (lower panel) distortion for a cluster with $y_{\rm sc}=10^{-4}$ at $z=0$ as a function of frequency and for different electron temperatures. The positive and negative parts of the intensity are drawn as solid and dotted lines, respectively. We also show the Kompaneets solution in solid black for reference.}
\label{fig:SD_shape}
\end{figure}


\begin{figure}
\includegraphics[width=\columnwidth]{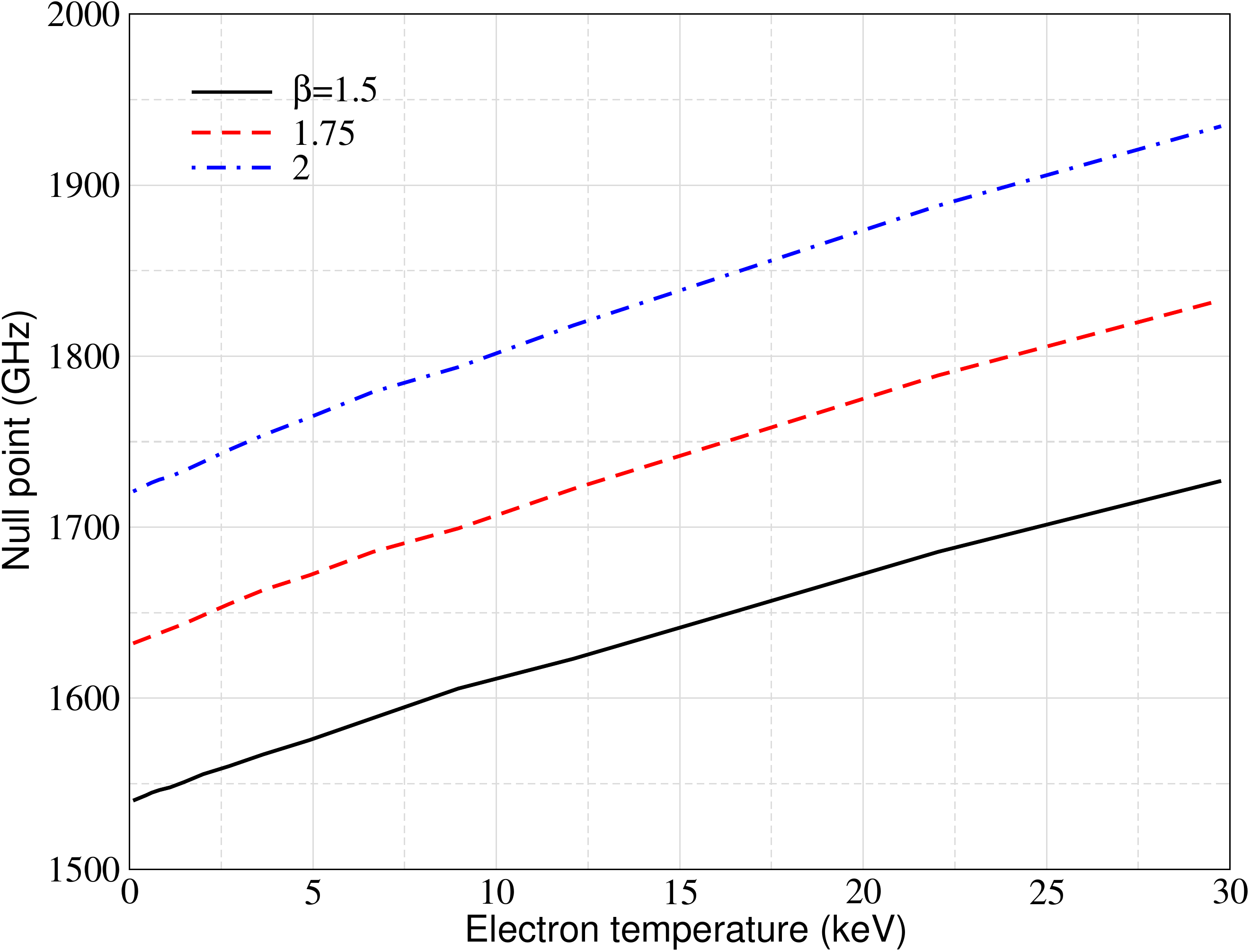}
\\[2mm]
\includegraphics[width=\columnwidth]{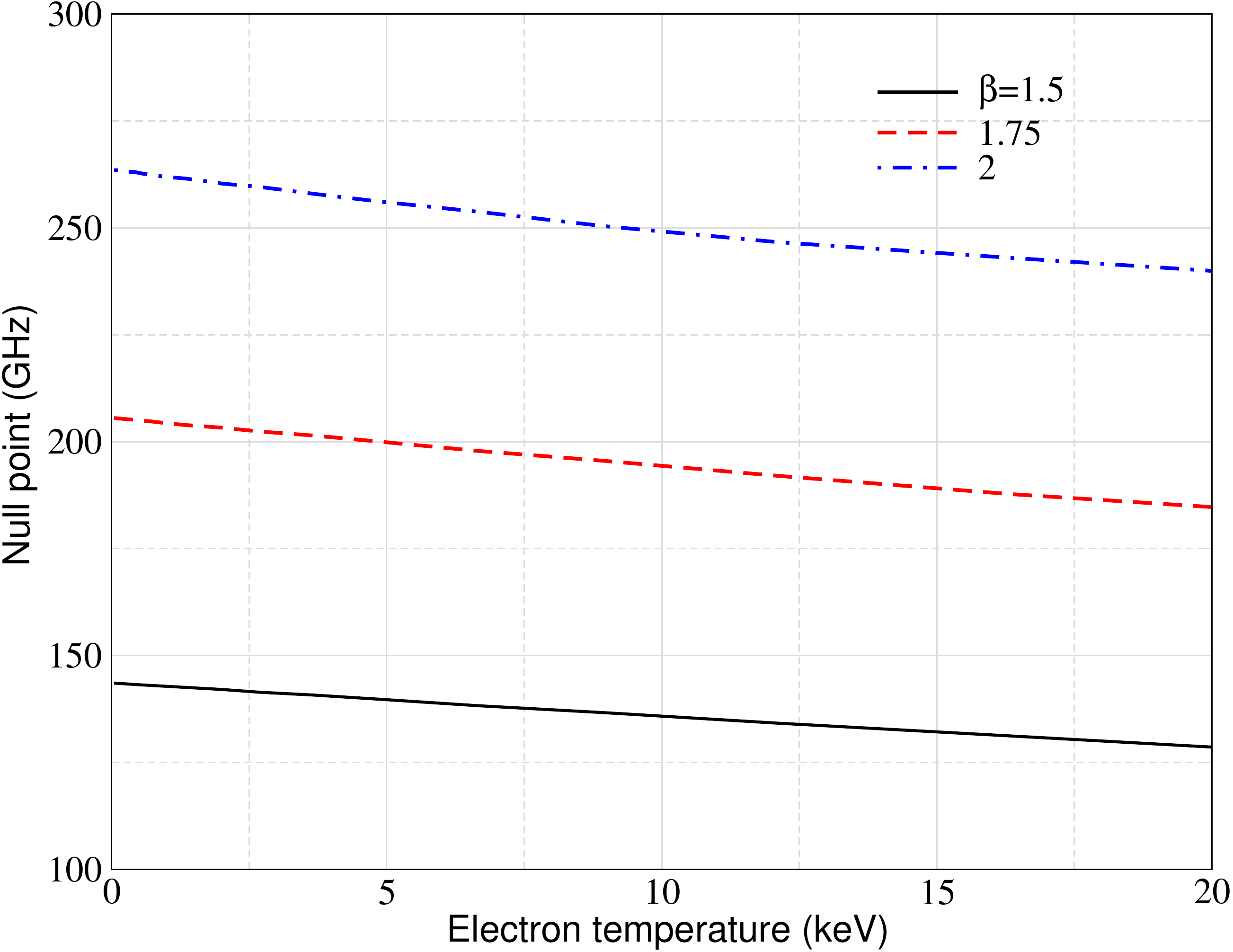}
\\[-0mm]
\caption{High (top panel) and low (lower panel) frequency null point as a function of the electron temperature and for varying values of $\beta$.}
\label{fig:null_point}
\end{figure}

\subsection{Spectral distortion with rSZ from individual object} 
\label{subsec:individual_object}
In this section, we compute the CIB spectral distortion signature for individual objects located at different redshifts and with varying electron temperature. For the scattering of the CMB, the $y$-distortion signature is independent of redshift, therefore, it is not possible to determine the redshift of the cluster using the spectral shape of $y$-distortion. But as has already been explained, the CIB keeps on evolving till today. Therefore, we expect that the spectral distortion signature imprinted at different redshifts will have different shapes. In principle, this provides the possibility to measure cluster redshifts.

\subsubsection{CIB scattering distortion at $z=0$} 
\label{subsec:individual_object_z0}
In Figure~\ref{fig:SD_shape}, we illustrate the scattered CIB signal as a function of frequency for a few different electron temperatures. We fixed the total scattering $y$-parameter ($y=\theta_{\rm e}\tau$) to $y_{\rm sc}=10^{-4}$. 
As opposed to the standard $y$-distortion \citep{Zeldovich1969}, one finds two nulls, where the intensity changes sign \citep{SHB2022}. For the CIB parameters chosen above, the nulls are found at $\nu_{\rm low}\approx 206\,{\rm GHz}$ and $\nu_{\rm high}\approx 1630\,{\rm GHz}$ in the non-relativistic (Kompaneets) limit. As the cluster temperature increases, the high-frequency null shifts towards higher frequency, while the low-frequency null decreases (see Fig.~\ref{fig:null_point}). 
For the chosen parameters, we find
\begin{subequations}
\begin{align}
\label{eq:fits_null}
\nu_{\rm low}(T_{\rm e})&\approx \,\,\,206\,{\rm GHz}\left[1-3.2\times 10^{-2} T_{\rm e,5}+1.7\times 10^{-3} T^2_{\rm e,5}\right]
\\
\nu_{\rm high}(T_{\rm e})&\approx 1630 \,{\rm GHz}
\left[1+2.4\times 10^{-2} T_{\rm e,5}-6.4\times 10^{-4} T_{\rm e,5}^2\right]
\end{align}
\end{subequations}
with $T_{\rm e,5}=
T_{\rm e}/[5\,{\rm keV}]$ to represent the positions of the nulls very well.
However, the positions of the nulls generally depends on $\beta$ (see Fig.~\ref{fig:null_point} for illustration) and $\gamma$, as well as the other CIB model parameters and details of the halo mass functions. 
The leading order $\beta$-dependence can be capture by 
\begin{subequations}
\begin{align}
\label{eq:fits_null_beta}
\nu'_{\rm low}(T_{\rm e}, \beta)&\approx 
\left[\frac{\beta}{1.75}\right]^{2.15}\,
\nu_{\rm low}(T_{\rm e})
\\
\nu'_{\rm high}(T_{\rm e}, \beta)&\approx \left[\frac{\beta}{1.75}\right]^{0.38}\,
\nu_{\rm high}(T_{\rm e})
\end{align}
\end{subequations}
Since one can quickly compute the resultant distortion using {\tt SZpack} we do not provide a more detailed discussion here.

{\it Why are two nulls present in the scattered CIB radiation, even if the scattering on average leads to an increase of the photon energy?}
For the final scattered signal, both the net energy transfer as well as the broadening of the radiation field matter. For parts of the incoming radiation field that have a negative derivative (like in the Wien tails of the CMB and CIB), the broadening only leads to an additional increase of the net up-scattering signature.
However, in parts of the incoming radiation field with positive slope (like in the low-frequency CMB and CIB domains), the balance between energy transfer and line-broadening becomes more delicate.

For flat photon spectra (like in the Rayleigh-Jeans tail of the CMB) the mean shift of the photon energy by $\Delta\nu/\nu\simeq 4 y_{\rm sc}$ dominates on average, leading to a decrement at $\nu\lesssim 217\,{\rm GHz}$ for the tSZ effect. However, when the photon distribution becomes sufficiently steep (like for the CIB spectrum at low frequencies), then also the line-broadening, $\Delta \nu/\nu\simeq \sqrt{2 y_{\rm sc}}$ can become noticeable. This leads to a net down-scattering of the incoming radiation, and hence an increase of the CIB signal in the direction of the cluster. 

\begin{figure}
\centering 
\includegraphics[width=0.95\columnwidth]{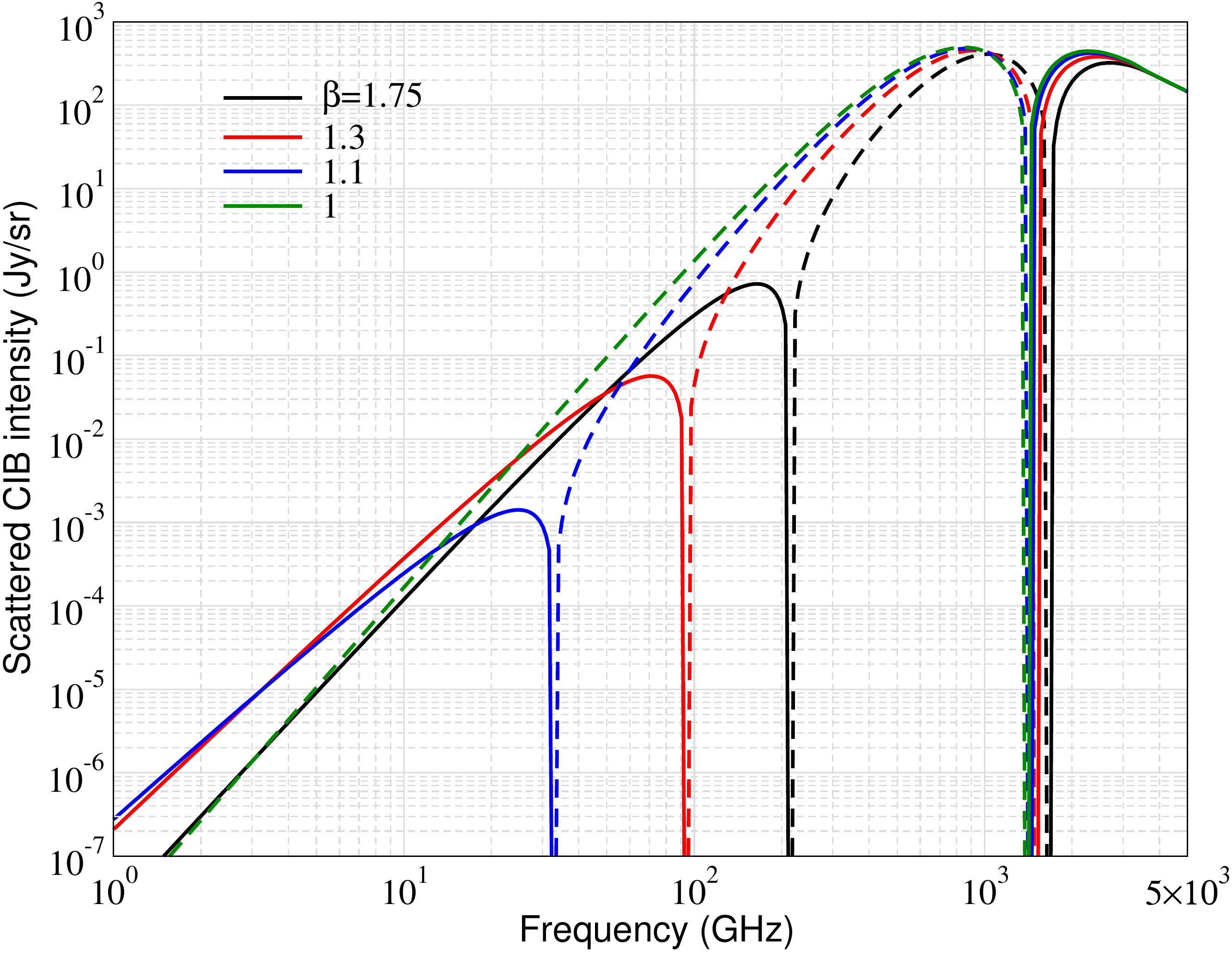}
\caption{Scattered CIB intensity for varying $\beta$ with $y_{\rm sc}=10^{-4}$ and at $z=0$. The position of the low-frequency and high-frequency null decreases with $\beta$. For $\beta=1$, the low-frequency null disappears.}
\label{fig:beta_dependence}
\end{figure}

For the chosen CIB parameters, we find $\beta\leq 1$ to remove the CIB null at low frequencies (see Fig.~\ref{fig:beta_dependence}). This is well below the observational result of $\beta\simeq 1.75$ at $z=0$, and thus is not expected to be very relevant in applications. The position of the low frequency null can be estimated by determining at which frequency $\Delta \Theta_{\rm sc}(\nu', z')$ vanishes. Assuming $x'\ll 1$, a limit that become increasingly accurate with decreasing $\beta$, one finds $x'_{\rm low}\approx (\beta^2+\beta-2)/(1+\beta)$, which implies the critical value $\beta_{\rm cr}= 1$, as confirmed in Fig.~\ref{fig:beta_dependence}.

\begin{figure}
\centering 
\includegraphics[width=\columnwidth]{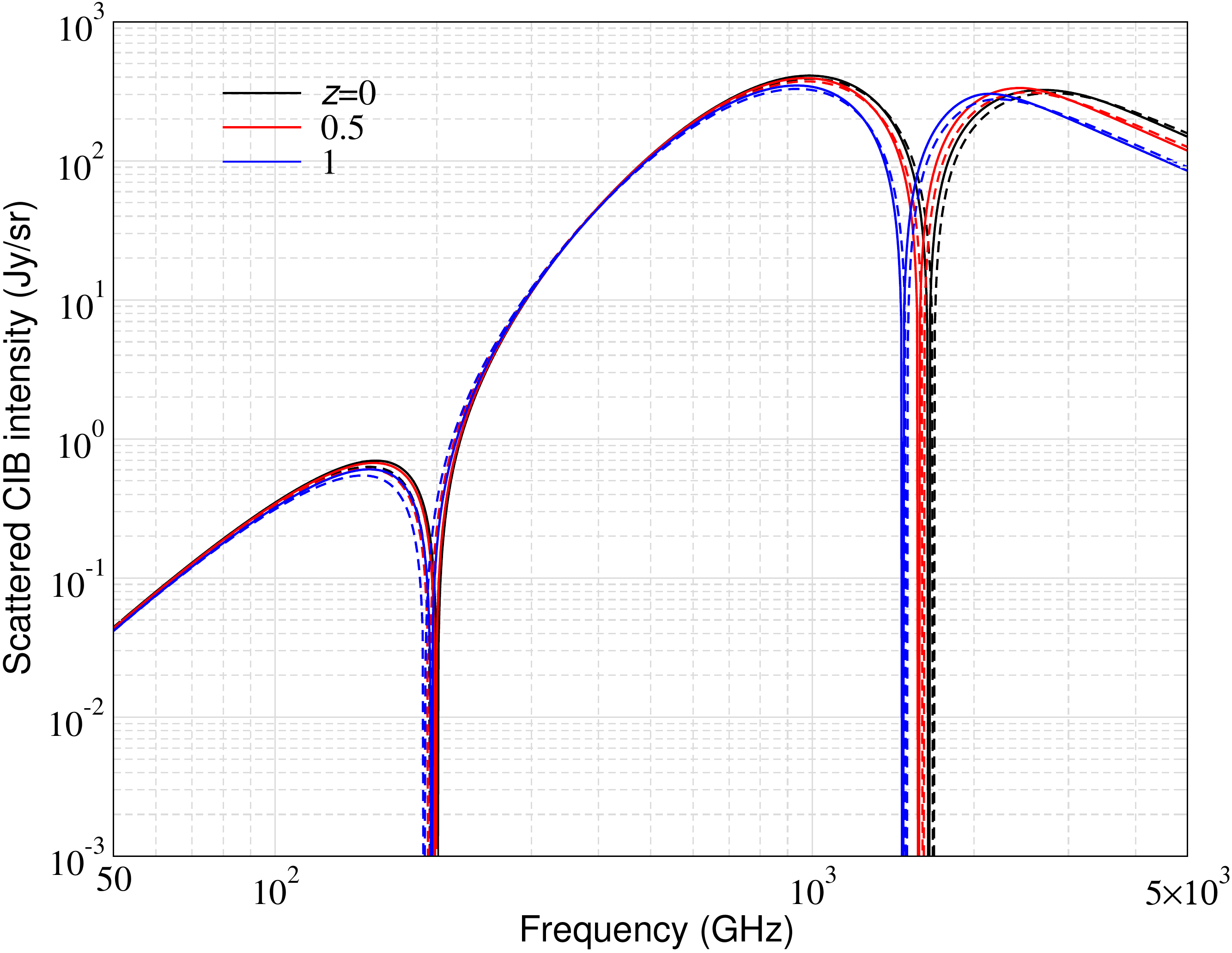}
\\[2mm]
\includegraphics[width=\columnwidth]{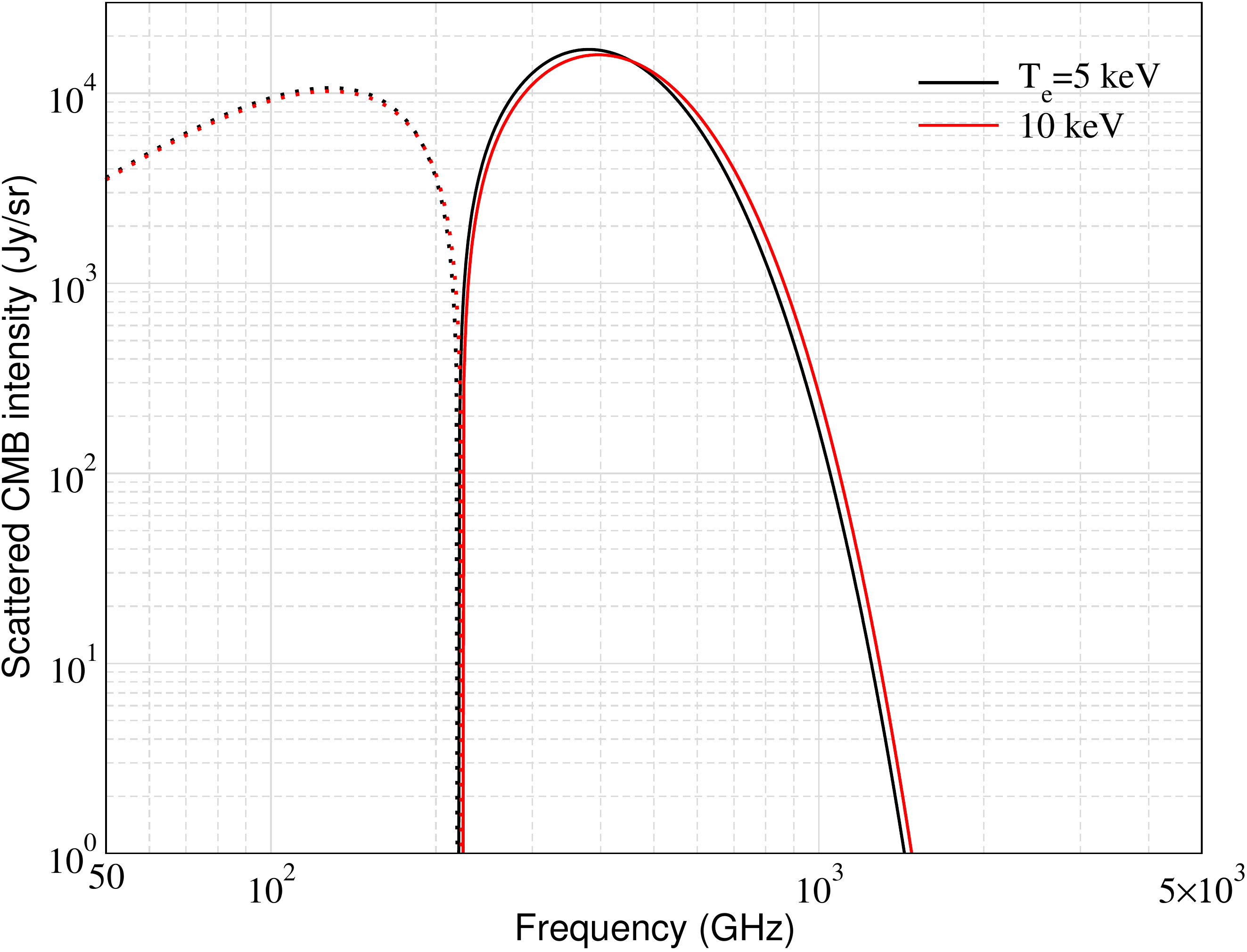}
\\
\caption{CIB (upper panel) and CMB (lower panel) distortions from individual galaxy clusters with $y_{\rm sc}=10^{-4}$, located at different redshifts and for two temperatures, 5~keV (solid line) and 10~keV (dashed lines). 
The dotted lines shows the negative part of intensity. 
For CMB, the scattered radiation is independent of redshift, while for the CIB a clear dependence is found.}
\label{fig:SD_single_object}
\end{figure}

\begin{figure}
\includegraphics[width=\columnwidth]{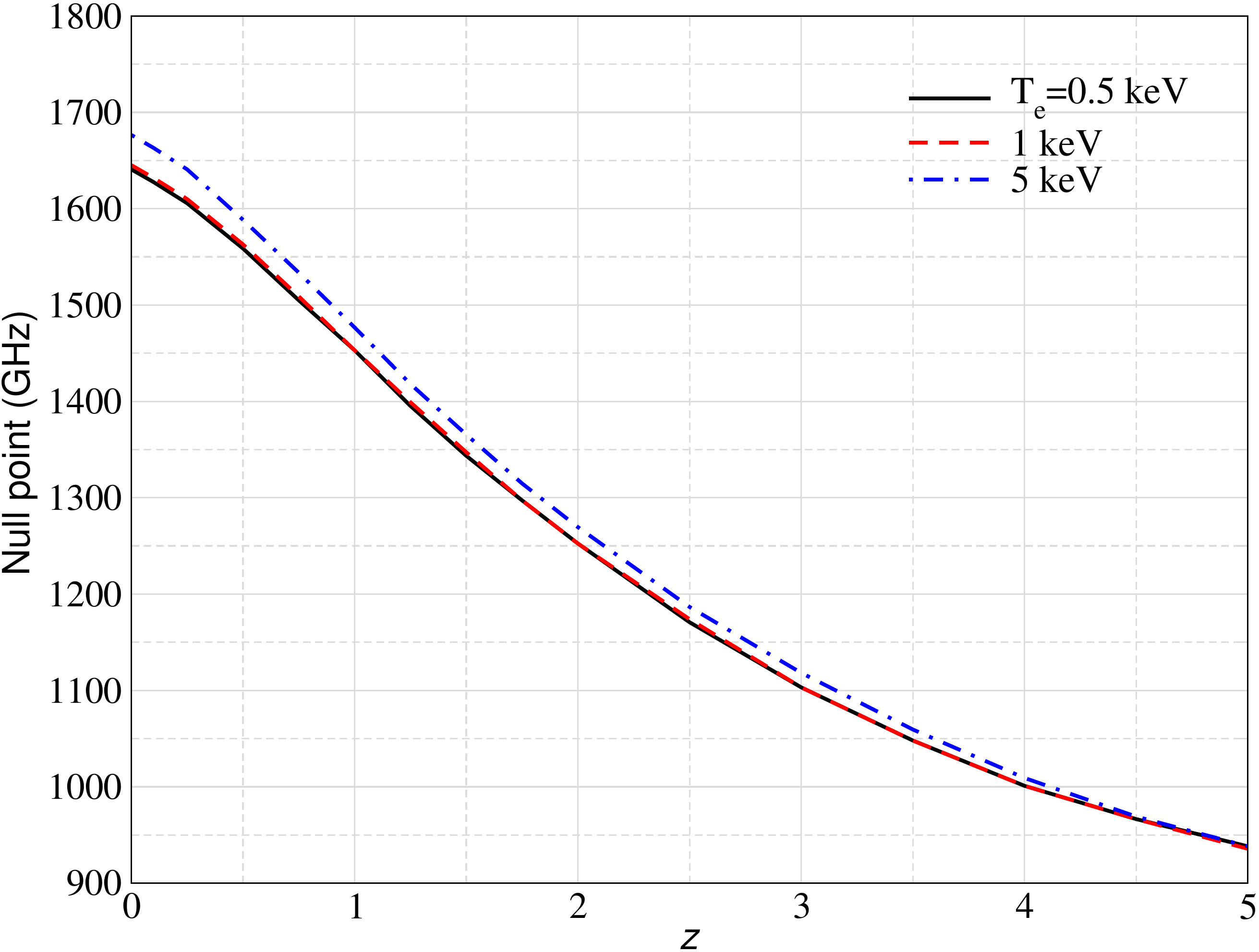}
\\[2mm]
\includegraphics[width=\columnwidth]{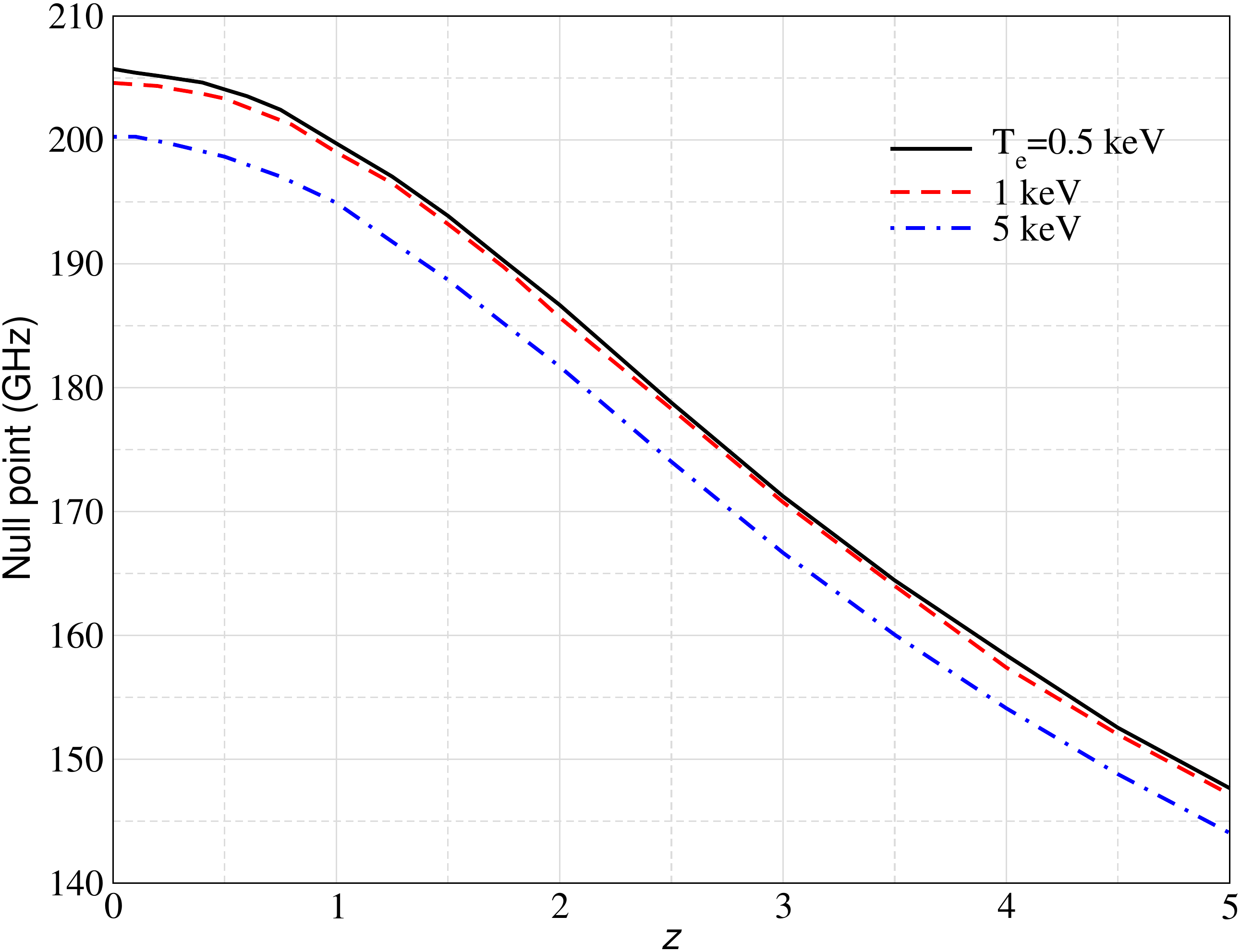}
\\[-0mm]
\caption{Redshift dependence of the high (upper panel) and low (lower panel) frequency null. In each figure we show three different temperature cases as annotated. The case with $T_{\rm e}=0.5\,{\rm keV}$ is very close to the Kompaneets solution (see Fig.~\ref{fig:null_point_z_approx}).}
\label{fig:null_point_z}
\end{figure}

\vspace{-3mm}
\subsubsection{CIB scattering distortion at varying redshift} 
\label{subsec:individual_object_zvar}
As already mentioned, we expect the scattered CIB signal to depend on the redshift of the cluster. 
In Fig.~\ref{fig:SD_single_object}, we illustrate this point for an individual object with $y_{\rm sc}=10^{-4}$ and temperature of the order of a few keV. 
We can observe both a change in the overall amplitude of the signal as well as a change in the frequency shape and position. At a fixed scattering $y$-parameter, this effect is {\it not} present for the tSZ effect (lower panel of Fig.~\ref{fig:SD_single_object}). By combining tSZ and scattered CIB measurements, one can thus in principle measure the redshift of the cluster. This may open the path for tomographic studies using the CIB spectral distortion signature from individual objects. 

However, as Fig.~\ref{fig:SD_single_object} also shows, since the shape of the scattered CIB also depends on the cluster temperature, a modeling of the relativistic corrections will be required to achieve high precision. This can be either done using theoretical temperature-mass relations \citep{Lee2020scalings}, X-ray proxies, or possibly through future rSZ measurements. To disentangle tSZ and scatter CIB contributions it is clear that high-frequency coverage will be very important.

In Fig.~\eqref{fig:null_point_z} we illustrate the redshift evolution of the low and high frequency nulls. With increasing redshift both move towards lower frequencies, with the high-frequency null changing more drastically. For observations towards a single cluster, one may hope to use this information in the future.

\begin{figure}
\includegraphics[width=\columnwidth]{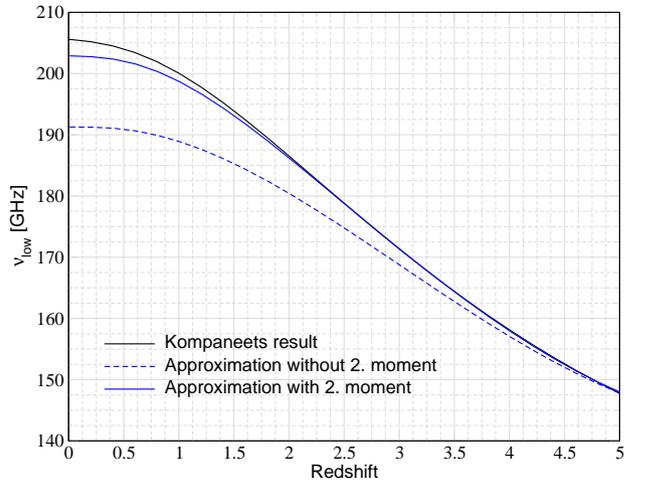}
\\[-3mm]
\caption{Redshift-dependence of the low-frequency null. The Kompaneets result is shown in comparison with the analytic approximation, Eq.~\eqref{eq:xnull_low_sol_M2}. At low redshifts the second moment term is clearly visible.}
\label{fig:null_point_z_approx}
\vspace{-3mm}
\end{figure}

\subsubsection{Analytic approximations for the low-frequency null}
\label{subsec:low_null_appr}
We have seen in Sect.~\ref{sec:ana_CIB} that Eq.~\eqref{eq:sol_ICIB_max} provides a very good approximation for the ambient CIB at different redshifts. If we neglect the correction from the second moment, we can obtain a simple estimate for the position of the null at the scattering redshift by solving
\begin{align}
\label{eq:xnull_low_condition}
Y(x^*)-2\beta\,G(x^*)+\beta(3+\beta)\approx 0.
\end{align}
This expression follows directly from Eq.~\eqref{eq:sol_ICIB_sc_op} after we apply the scattering operator to the analytic approximation. It is easy to numerically solve this expression. We find 
\begin{align}
\label{eq:xnull_low_sol}
x^*_{\rm low}(\beta)
&\approx 
0.957+ 1.964\,\zeta
-0.435\,\zeta^2
+0.324\,\zeta^3
-0.236\,\zeta^4
\end{align}
with $\zeta=\beta/1.75-1$ to be extremely accurate at $1.25\leq\beta\leq 2.5$. 
%
%
Because this solution is for $x^*_{\rm low}(\beta)=x(z_{\rm sc})\,\xi^*(z_{\rm sc})$, we then have
\begin{align}
\label{eq:nunull_low_sol}
\nu_{\rm low}(z_{\rm sc},\beta)
&\approx
\frac{191.2\,{\rm GHz}}{(1+z_{\rm sc})^{1-\alpha}}
\left[\frac{T_0}{24.4\,{\rm K}}\right]
\,
\left[\frac{x^*_{\rm low}(\beta)}{0.957}\right]
\,
\left[\frac{\xi^*(z_{\rm sc})}{2.544}\right]^{-1}
\end{align}
today. Using Eq.~\eqref{eq:xifit}, we can then also compute the null for higher redshifts. This implies that the null decreases with redshift. 

When we compare the result for the null with the one obtained from the exact computation, we realize that the analytic expression slightly underestimates the position of the low-frequency null. The reason is that we neglected the second moment term in Eq.~\eqref{eq:sol_ICIB_max}. Using the scattering operator $\hat{\mathcal{D}}_{x^*}=x^* \partial_{x^*} (x^*)^4 \partial_{x^*} (x^*)^{-3}$, and applying it to the analytic solution for the CIB, after some algebra we find the modified condition
\begin{align}
&Y(x^*)-2\beta\,G(x^*)+\beta(3+\beta)
\nonumber
\\
\nonumber
&\qquad+\frac{\frac{M_2}{2}\,\left[ (x^*)^2\partial^2_{x^*} Y^*(x^*)
+2\left[2+\beta - G(x^*)\right] 
x^*\partial_{x^*} Y^*(x^*)\right]
}{1+\frac{M_2}{2}\,Y^*(x^*)} \approx 0.
\end{align}
for the low-frequency null. Perturbing around the solution from above and accounting for leading order corrections, we then find
\begin{align}
\label{eq:xnull_low_sol_M2}
\tilde{x}^*_{\rm low}(\beta, M_2)
&\approx x^*_{\rm low}(\beta)+1.497\,M_2
\end{align}
For $z=0$, one has $M_2\approx 0.039$ and hence $\tilde{x}^*_{\rm low}\approx 1.015$. Inserting this into Eq.~\eqref{eq:nunull_low_sol} instead of $x^*_{\rm low}$ yield $\nu_{\rm low}(0, 1.75)\approx 203.2\,{\rm GHz}$. This is already very close to the full numerical result. A direct comparison is shown in Fig.~\ref{fig:null_point_z_approx}.

\subsection{Mean intensity of spectral distortion due to all objects }
\label{subsec:meany_intercluster}
In this section, we stack the individual objects to obtain the globally-averaged CIB distortion signal.
The computation of mean distortion signal from CIB photons is very similar to that of the average CMB $y$-distortion \citep{Hill2015}. The distortion to the infrared background is proportional to the $y$-parameter of electrons inside the dark matter halos. The average scattering $y$-parameter is given by \citep{Hill2015, SHB2022},
\begin{equation}
    \langle y \rangle=\int_0^{z_{\rm max}}\frac{{\rm d}y}{{\rm d}z}\,{\rm d}z 
    \;\,\text{with}\;\,
    \frac{{\rm d}y}{{\rm d}z}=\frac{c\chi^2}{H(z)}\,\int_{M_{{\rm min}}}^{M_{\rm max}}{\rm d}M \,\frac{{\rm d}N}{{\rm d}M}\,y(M,z),
    \label{eq:y_param_def}
\end{equation}
where the average halo $y$-parameter is (see Appendix~\ref{app:y_cos} for additional clarifications)
\begin{equation}
\label{eq:y_M_def}
    y(M,z)=\frac{\sigma_T}{m_{\rm e} c^2}\frac{4\pi r_{\Delta}^3(z)}{d_{\rm A}(z)^2}\int {\rm d}x \,x^2 P_{\rm e}(x),
\end{equation}
with $x=r/r_{\Delta}$, $r_{\Delta}(z)=\left[\frac{3M}{4\pi \Delta\rho_{c}(z)}\right]^{1/3}$ and $P_{\rm e}(x)$ is the electron pressure. Here, $\chi(z)$ is the comoving distance to $z$ and $d_A(z)=a \,\chi(z)$ is the angular diameter distance. We use the overdensity $\Delta=200$ throughout our calculations. 

Using Eq.~\eqref{eq:sol_ICIB_sc_z0}, the intensity of distorted signal is then given by,
\begin{align}
\label{eq:sol_ICIB_sc_meany}
    \Delta I^{\rm CIB}_{\nu, 0}&\approx 
    \int_0^{z_{\rm max}} {\rm d}z_{\rm sc}\,\frac{{\rm d}y}{ {\rm d}z_{\rm sc}}
    \int^{z_{\rm max}}_{z_{\rm sc}} 
    \frac{c\,{\rm d} z'}{H(z')}
    \frac{a'\bar{L}(z')}{4\pi}
    \,\Delta \Theta_{\rm sc}(\nu/a', z')
\end{align}
This expression relies on the fact that the spectral distortion shape is independent of the gas temperature/mass of the halo. In the Kompaneets approach, this is indeed possible, as $\Delta \Theta_{\rm sc}(\nu/a', z')$ in Eq.~\eqref{eq:sol_ICIB_sc_z0} is mass-independent. One can then integrate over  mass before computing the intensity of the distorted signal at a given redshift. 

However, we wish to take into account relativistic temperature corrections, which are a function of the temperature (or mass) of halo. In this case, Eq.~\eqref{eq:sol_ICIB_sc_meany} has to be modified to,
\begin{align}
\label{eq:sol_ICIB_sc_meany_rsz}
    \Delta I^{\rm inter}_{\nu, 0}&\approx \int_0^{z_{\rm max}}{\rm d}z_{\rm sc}\int_{M_{\rm min}}^{M_{\rm max}} {\rm d}M_{\rm sc}\frac{{\rm d}y}{ {\rm d}z_{\rm sc}{\rm d}M_{\rm sc}}  \nonumber \\
    & \qquad
    \times \int^{z_{\rm max}}_{z_{\rm sc}} \frac{c\,{\rm d} z'}{H(z')}
    \,\frac{a'\bar{L}(z')}{4\pi}
    \,\Delta \Theta_{\rm sc}(\nu/a', z',M_{\rm sc}),
\end{align}
where,
\begin{equation}
    \frac{{\rm d}y}{{\rm d}z{\rm d}M}=\frac{c\chi^2}{H(z)}\, \frac{{\rm d}N}{{\rm d}M}\,y(M,z),
\end{equation}
and $ \Delta \Theta_{\rm sc}(\nu/a', z',M_{\rm sc})$ is the spectral distortion shape computed using the kernel method described in Sec.~\ref{sec:SD_shape_Kernel}. 

\begin{figure}
\centering 
\includegraphics[width=0.95\columnwidth]{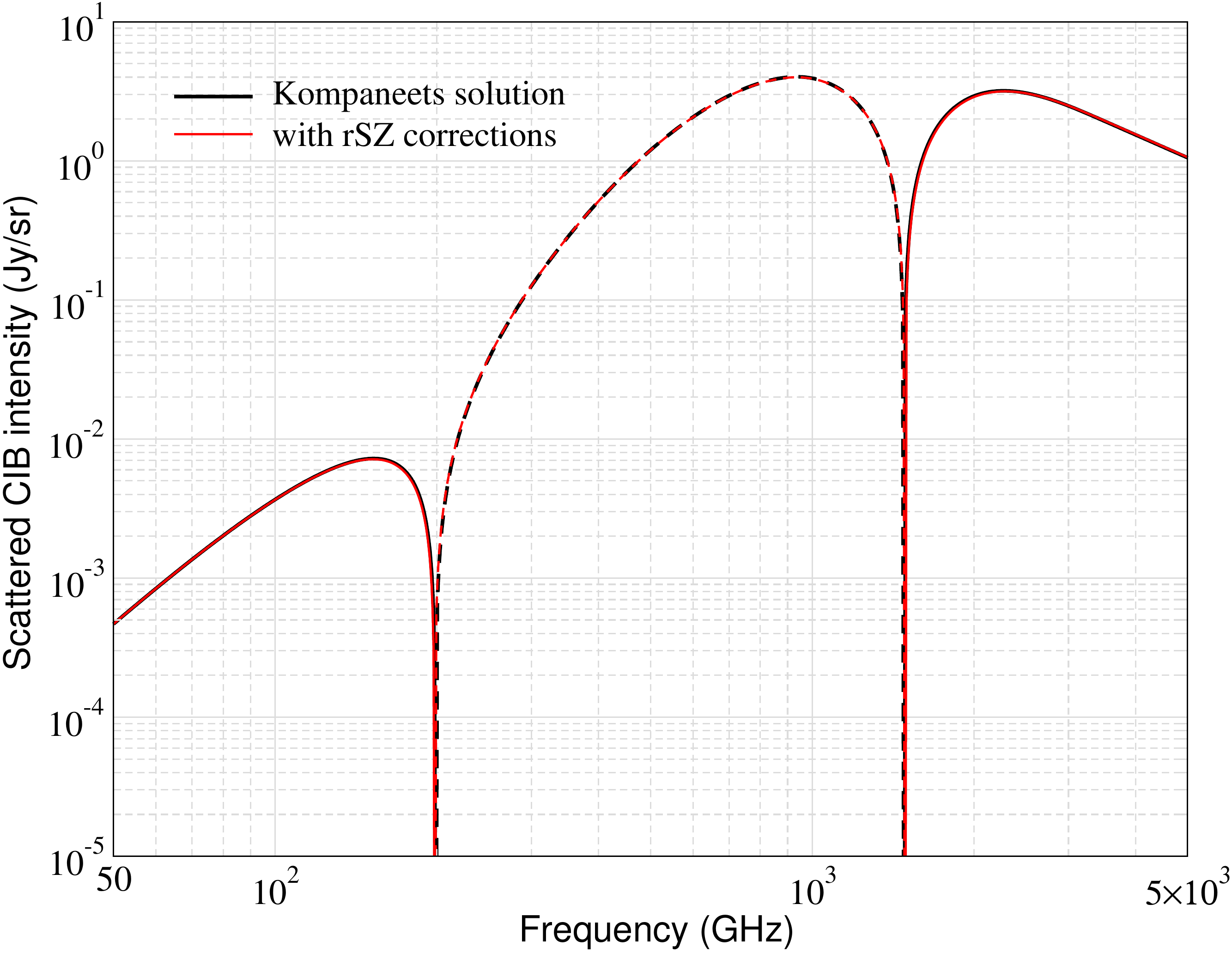}
\caption{Intensity of spectral distortion signal from intercluster  scattering of CIB photons assuming non-relativistic scattering (black line) and including relativistic effects (dashed red).}
\label{fig:SD_rsz}
\end{figure}

In Fig. \ref{fig:SD_rsz}, we plot the globally-averaged CIB distortion signal today ignoring and taking into account relativistic effects, respectively. 
We use the $M$-$T$ relations of \citep{Lee2020scalings} to include relativistic corrections.
The first aspect to notice is that the average cCIB has a shape that is very similar to that for a single cluster. However, the amplitude is suppressed by about two orders of magnitude. This is consistent with the fact that the average $y$-parameter is $\langle y\rangle \simeq 10^{-6}$ \citep{Hill2015} as opposed to $y_{\rm sc}\simeq 10^{-4}$ used in the single cluster illustrations (see Fig.~\ref{fig:SD_shape} and \ref{fig:SD_single_object}).

The second point to note is that the positions of the nulls are a little lower than for a single cluster scattering at $z=0$. For the all-sky average cCIB, we find $\nu_{\rm low}\approx 199\,{\rm GHz}$ and $\nu_{\rm high}\approx 1480\,{\rm GHz}$, which is consistent with the result of \citet{SHB2022}. The difference relative to the single cluster is due to the overall redshift dependence of the cCIB signal and the fact that most of the signal originates from $z\simeq 1$, hence having lower crossover frequencies (see Fig~\ref{fig:SD_single_object}).

Finally, the differences in the cumulative CIB scattering signal from relativistic effects are not very large (see Fig.~\ref{fig:SD_rsz}). we find $\nu_{\rm low}\approx 198\,{\rm GHz}$ and $\nu_{\rm high}\approx 1485\,{\rm GHz}$ This is because on average there are very few heavy halos (which have high temperatures) due to steep decline of halo mass function for high overdensity. This is consistent with the results of \cite{Hill2015} who obtained the mean temperature of gas to be about 1.3 keV for which one does not see much difference from non-relativistic solution (Kompaneets solution in Fig. \ref{fig:SD_shape}) except at the very high frequency tail. We conclude that on average relativistic temperature corrections to the CIB signal will be hard to extract, however, for studies on a cluster-by-cluster basis they should be included.

\vspace{-0mm}
\section{Scattering within a halo}
\label{sec:cluster_scattering}

Up to this point, we have assumed that the CIB photons escape their parent halo un-scattered to give rise to CIB background that we see today. They scatter predominantly with electrons inside massive halo which form only at $z\lesssim 1$. We name this process "intercluster" scattering.
However, in addition there will be an "intracluster" contribution, which is when the CIB photons scatter with the electrons inside the parent halo. In this section, we compute the scattered CIB contribution of each halo due to intracluster scattering. 

\vspace{-0mm}
\subsection{Kinetic equation for scattering inside the cluster}
To compute the intracluster scattering contribution, we start with 
the radiative transfer equation for photon in the anisotropic cluster medium. We shall assume that the cluster is spherically-symmetric and that the single-scattering approximation can be made. For computing the scattered radiation, the Hubble-expansion can be neglected and only local physics need to be considered, as already used above. We then have the transfer equation
\begin{equation}
    \frac{\partial I_\nu(r,\hat{\vek{\gamma}})}{\partial \tau}\approx C[I_\nu(r,\hat{\vek{\gamma}})],
\end{equation}
where $r$ is the distance with respect to the cluster center, $\hat{\vek{\gamma}}$ defines the line of sight direction and $\tau$ is the corresponding optical depth along the photon path. 

To make progress, we have to evaluate the collision term. We assume that is only is related to electron scattering and that the unscattered radiation field can be locally computed just by integrating all sources.
The expression for $C[I_\nu]$ to linear order in temperature is given by \citep[e.g.,][]{CDK2014},
 \begin{align}
 \label{eq:intracluster}
&C[I_\nu]=\frac{3}{16\pi}\int {\rm d}^2\hat{\vek{\gamma}}' (1+\mu^2)[I_\nu(r,\hat{\vek{\gamma}}')-I_\nu(r,\hat{\vek{\gamma}})] 
\nonumber\\
&\qquad\qquad
+\frac{3 \theta_{\rm e}}{16\pi} \int {\rm d}^2\hat{\vek{\gamma}}' [2-4\mu-6\mu^2+4\mu^3] \, I_\nu(r,\hat{\vek{\gamma}}')] 
\\ \nonumber
&\qquad\qquad\quad
+\frac{3 \theta_{\rm e}}{16\pi}\,\nu \frac{\partial}{\partial \nu}\nu^4\frac{\partial}{\partial \nu}\int {\rm d}^2\hat{\vek{\gamma}}'(1+\mu^2)(1-\mu)\,
\frac{I_\nu(r,\hat{\vek{\gamma}}')}{\nu^3},
 \end{align}
 where $I_\nu(r,\gamma')$ is the radiation at location $r$ coming from direction $\hat{\vek{\gamma}}'$ and $\mu=\hat{\vek{\gamma}}\cdot\hat{\vek{\gamma}}'$ is the scattering angle. 
 The first line is just the Thomson scattering term without energy exchange between electron and photon, while 
 %
 the second term describes the correction to Thomson scattering cross-section to first order in electron temperature. These term do not alter the spectral shape of the incoming CIB signal, and we therefore omit them here.\footnote{If there is any effect from these terms, this will lead to a slight re-weighting of the CIB flux from each halo. However, photon number conservation suggests these terms to drop out once averaged over the parent halo.}
 
 The last term is the energy exchange term, which will give intracluster contribution to CIB spectral distortions. The scattering operator is again the Kompaneets operator, however, the coupling extends from monopole to octupole of the local radiation field \citep{CDK2014}. We shall neglect relativistic corrections to the scattered signal, which can be included using the anisotropic scattering kernel \citep[e.g.,][]{Chluba2014mSZ}.
 
 Our task is now to compute the scattering integrals. We can decompose the radiation at location $r$ into multipoles as,
 \begin{equation}
     I_\nu(r,\mu)=\sum_{\ell} P_{\ell}(\mu)\,I_{\nu,{\ell}}(r),
 \end{equation}
 where $P_\ell(\mu)$ are Legendre polynomials. Due to the symmetry of the scattering process, no terms depending on $m$ are needed. The expression for $I_{\nu,\ell}(r)$ can be written as,
 \begin{equation}
     I_{\nu,\ell}(r)=\frac{2\ell+1}{4\pi}\int {\rm d}^2 \hat{\vek{\gamma}'}\, P_\ell(\mu)\,I_\nu(r,\hat{\vek{\gamma}}').
     \label{eq:multipole_definition}
 \end{equation}
 Plugging this into Eq.~\eqref{eq:intracluster}, and neglecting the Thomson terms, we have \citep{CDK2014},
 \begin{align}
 \label{eq:intracluster_multipole}
&C[I_\nu]\approx 
\theta_{\rm e} \hat{\mathcal{D}}_\nu\left[I_{\nu, 0}(r)-\frac{2}{5}I_{\nu, 1}(r)+\frac{1}{10}I_{\nu, 2}(r)-\frac{3}{70}I_{\nu, 3}(r)\right].
 \end{align}
with the scattering operator $\hat{\mathcal{D}}_\nu=\nu \partial_\nu \nu^4 \partial_\nu \,\nu^{-3}$.
We next need to specify the terms $I_{\nu, \ell}(r)$, for which we need to model the intracluster CIB luminosity. However, before, we treat the scattering of light from the central galaxy first and then treat the scattering from the sub-halo sources as a special case.

\subsection{Scattering of light from the central galaxy}
The effect of scattering on the light from the central galaxy can be computed quite easily. For this we write the central luminosity, $L_{\nu}^{\rm c}(M,z)=N^{\rm c}(M,z)\,L_{\nu}^{\rm gal}(M,z)$, as $L_{\nu}^{\rm c}(M,z)=\bar{L}^{\rm c}(M,z)\,\Theta_\nu(z)$. The source of photons is located at the center of the cluster, such that the spectral intensity at a distance $r$ from the center simply is $I^{\rm c}_\nu(r)={\rm e}^{-\tau_{\rm c}(r)}\,I_{\nu}^{\rm c}(M,z)$, where $I_{\nu}^{\rm c}(M,z)=\bar{I}^{\rm c}(M,z)\,\Theta_\nu(z)$ is the corresponding intensity of the source, which we determine below. Here $\tau_{\rm c}(r)$ is the total optical depth from the center to the location at distance $r$. The factor ${\rm e}^{-\tau_{\rm c}(r)}$ takes into account the scattering of photons out of the line of sight while they travel toward the scatterer at $r$. Since we work in the optically-thin limit, we drop this factor in the evaluation of the collision term. 

Considering a line of sight through the cluster with impact parameter $b$ from the center, and using $s$ to denote the location along the line of sight then we have $r(b,s)=\sqrt{s^2+b^2}$ (with $s=0$ when $r=b$ and $s>0$ away from the observer). The angle with respect to the line of sight is then given by $\mu(b,s)=s/r$. We can then directly carry out all the angular integrals, which results in 
\begin{align}
\label{eq:DI_into}
\frac{\partial I_\nu(s,b)}{\partial s}&\approx
\frac{3}{8}\,\frac{{\rm d}y}{{\rm d}s}\,(1+\mu^2)(1-\mu)
\,\hat{\mathcal{D}}_\nu\,I_{\nu}^{\rm c}(M,z),
\end{align}
where it is understood that $\mu=\mu(s,b)$ and $r=r(s,b)$. Here, the differential ${\rm d}y/{\rm d}s=\theta_{\rm e}{\rm d}\tau/{\rm d}s=\theta_{\rm e}\,N_{\rm e}(r)\,\sigma_{\rm T}$ defines the $y$-parameter along the line of sight:
\begin{align}
y(b, s_{\rm min}, s_{\rm max})=\int^{s_{\rm max}}_{s_{\rm min}} \theta_{\rm e}\,N_{\rm e}\left(\sqrt{s^2+b^2}\right)\,\sigma_{\rm T}\,{\rm d}s.
\end{align}
For convenience, we define the total $y$-parameter along the line of sight with impact parameter $b$ as $y(b)=y(b, -R, R)$, where $R$ determines the size of the system. Integrating over all lines of sight and dividing by the projected area gives the volume average $y$-parameter. 

From Eq.~\eqref{eq:DI_into}, we can in principle compute the scattered CIB signal along each line of sight through the cluster. 
%
%
However, here we a mainly interested in the average signal over the cluster. 
To compute the result, we use cylindrical coordinates along the direction through the cluster center. The averaged scattered CIB signal in the direction of the cluster is then given by the integral 
\begin{align}
\Delta \bar{I}_\nu^{\rm c}=\int \frac{{\rm d}s \,b {\rm d}b}{\pi R^2} \,{\rm d}\phi \, \partial_s I_\nu(s,b), 
\end{align}
which assumes that the size of the system is finite within the radius $R$. To carry out the integral, we can think of the average as a volume integral around the center of the cluster, ${\rm d}s \,b {\rm d}b \,{\rm d}\phi\rightarrow r^2{\rm d}r \,{\rm d}\mu_{\rm r} \,{\rm d}\phi$, where $\mu_r$ is the direction cosine of the location of the electron at $(s,b)$ in the new system. Since in this situation we have the $\mu\equiv \mu_r$ for the scattering angle, all the integrals can be performed independently.
The integral over $\phi$ simply yields a factor of $2\pi$.
Carrying out the integration over $\mu_{\rm r}$ then yields
\begin{align}
\label{eq:central_sol}
\Delta \bar{I}_{\nu}^{\rm c}
&\approx
\frac{2}{R^2}\left(\int r^2 {\rm d}r
\,\theta_{\rm e}\,N_{\rm e}\left(r\right)\,\sigma_{\rm T}\right)
\,\hat{\mathcal{D}}_\nu\,I_{\nu}^{\rm c}(M,z)
\nonumber\\
&=\frac{y_{V}}{2}\,\hat{\mathcal{D}}_\nu\,I_{\nu}^{\rm c}(M,z),
\end{align}
where $y_{V}=\int \theta_{\rm e}\,N_{\rm e}\left(r\right)\,\sigma_{\rm T}\,{\rm d}V/[\pi R^2]$ is the volume average of the $y$-parameter. 
The cCIB contribution from the central galaxy thus has an amplitude $1/2$ of what an external radiation field with intensity $I_{\nu}^{\rm c}(M,z)$ would yield. Assuming a constant density scattering sphere this makes perfect sense because the photons transverse only half of the cluster.
Since the contributions of the central galaxies to the CIB dominate the total CIB, we therefore expect another contribution comparable to $1/2$ of the intercluster scattering signal. However, due to the non-vanishing average distance between clusters of a given mass $M$, for the contributions from the intracluster scattering the signal is slightly enhanced by the lack of redshifting before the scattering event. We will illustrate this point below.

\subsection{Scattering of light from the sub-halo galaxies}
To compute the contributions from the sub-halo galaxies, we introduce the spherically-symmetric emission profile\footnote{This has units ${\rm Jy\, sr^{-1}\,Mpc^{-1}}$ and will be determined below.}, $j_{\nu}^{\rm s}(r, M,z)$, of isotropic emitters. To compute the intensity in a given direction $\hat{\vek{\gamma}}'$ we need to integrate along this direction from the location $\vek{r}(s,b)$ of the scattering electron to infinity:
\begin{equation}
\label{eq:halo_I_gammap}
I^{\rm s}_{\nu}(s, b, \mu, M,z)=\int^\infty_0 {\rm d}s^* j_{\nu}^{\rm s}(r^*, M,z),
\end{equation}
with $r=\sqrt{s^2+b^2}$ and $r^*=\sqrt{r^2+2 \mu s^*+(s^*)^2}$, where $s^*$ parameterizes the integral along the direction $\hat{\vek{\gamma}}'$. One has $r^*=r$ at $s^*=0$, which marks the starting point of the integration.

If we then compute the Legendre transforms of the corresponding cumulative radiation field around the location $\vek{r}$, one can show that \citep[see arguments in Sect.~4 of][for the $y$-parameter]{CDK2014}
\begin{subequations}
\begin{align}
\label{eq:multipole_definition_halo}
I^{\rm s}_{\nu,\ell}(s, b, M,z)
&=P_\ell(s/r)\,I_{\nu, \ell}^{\rm s, 0}(s, M,z)
\\
\label{eq:multipole_definition_special}
I_{\nu, \ell}^{\rm s, 0}(s, M,z)
&=
\frac{2\ell+1}{4\pi}\int {\rm d}^2 \hat{\vek{\gamma}'}\, P_\ell(\mu)\,I^{\rm s}_\nu(s, 0, \mu, M,z).
\end{align}
\end{subequations}
Using the collision term, Eq.~\eqref{eq:intracluster_multipole}, we then have
 \begin{align}
 \label{eq:intracluster_multipole_halo_fin}
\frac{\partial I_\nu(s, b)}{\partial s}&\approx
\frac{{\rm d}y}{{\rm d}s}\,\hat{\mathcal{D}}_\nu
\left[I^{\rm s}_{\nu, 0}-\frac{2}{5}I^{\rm s}_{\nu, 1}+\frac{1}{10}I^{\rm s}_{\nu, 2}-\frac{3}{70}I^{\rm s}_{\nu, 3}\right],
 \end{align}
where all $I^{\rm s}_{\nu, \ell}=I^{\rm s}_{\nu, \ell}(s, b, M,z)$.
Although one can study the detailed spatial structure of the scattered radiation field using Eq.~\eqref{eq:intracluster_multipole_halo_fin}, we are again mainly interested in the cluster-averaged signal. For this we convert the 
integral into a volume average. Adapting the arguments of\footnote{Here, $I^{\rm s}_{\nu, \ell}$ plays the role of $y_\ell$ in \citet{CDK2014} and therefore all averages of $I^{\rm s}_{\nu, \ell}$ over $b{\rm d}b\,{\rm d}s$ will vanish unless $\ell=0$.} \citet[][Sect.~4.2]{CDK2014}, we can directly deduce
\begin{subequations}
\label{eq:halo_sol}
\begin{align}
\Delta \bar{I}_{\nu}^{\rm s}
&=y_{V} \,\hat{\mathcal{D}}_\nu\left<I^{\rm s}_\nu(M,z)\right>_y
\\
\left<I^{\rm s}_\nu(M,z)\right>_y&=
\frac{\int\int
\frac{j_{\nu}^{\rm s}(r, M,z)\,P_{\rm e}(r')}{4\pi|\vek{r}-\vek{r}'|^2}\,{\rm d}V \,{\rm d}V'}
{\int P_{\rm e}(r')\,{\rm d}V'},
\end{align}
\end{subequations}
where we now identified the $y$-parameter with the integrated electron pressure.
Together with Eq.~\eqref{eq:central_sol} for the contribution from the central galaxy, this allows us to give the average CIB intensity contribution from each halo, as we explain next.

\subsection{Total intracluster scattering contribution}
With the solutions Eq.~\eqref{eq:central_sol} and \eqref{eq:halo_sol}, we can now write the total intracluster scattering contribution to the average intensity in the direction of the cluster. With this result, we will be able to generalize the computation of the average cCIB including this contribution.

Adding Eq.~\eqref{eq:central_sol} and \eqref{eq:halo_sol}, we obtain the total intercluster scattering contribution to the CIB signal as
\begin{align}
\Delta \bar{I}_{\nu}^{\rm h}
\nonumber
&\approx
\frac{y_{V}}{2}\,\hat{\mathcal{D}}_\nu \,I_{\nu}^{\rm c}(M,z)
+y_{V}\,\hat{\mathcal{D}}_\nu\left<I^{\rm s}_\nu(M,z)\right>_y.
\end{align}
How is this related to the observed average distortion for a cluster at some redshift. From Eq.~\eqref{eq:Iav_initial} we know that the observed average unscattered intensity of the cluster is related to a volume-average of the emissivity. At cosmological distances this then is
\begin{align}
\bar{I}^{\rm h}_\nu 
&=\frac{1}{4\pi\,\chi^2\,a^3 }\,\frac{a L^{\rm h}_\nu(M,z)}{4\pi}=
\frac{1}{4\pi\,\chi^2\,a^3 } \int \tilde{j}_{\nu}^{\rm h}(r, M,z)\,{\rm d}\tilde{V}
\nonumber\\
&=
\frac{1}{4\pi\,\chi^2\,a^2 } \int j_{\nu}^{\rm h}(r, M,z)\,{\rm d} V
=
\frac{1}{4\pi\,d^2_{\rm A}} \int j_{\nu}^{\rm h}(r, M,z)\,{\rm d} V
\end{align}
at the halo's location at $z$. In the last step we related to physical quantities of the local system. In the simplified cylindrical geometry that we used, the equivalent of the averaged halo intensity is 
\begin{align}
\bar{I}^{\rm h, cyl}_\nu(M,z)
&=
\frac{1}{\pi R^2}\,\int j_{\nu}^{\rm h}(r, M,z)\,{\rm d}V, 
\end{align}
which we have to replace in our solution. We therefore divide the sub-halo term by this expression and then have 
\begin{subequations}
\label{eq:halo_sol_form}
\begin{align}
\Delta \bar{I}_{\nu}^{\rm h}&\approx
\frac{y_{V}}{2}\,\hat{\mathcal{D}}_\nu\,I_{\nu}^{\rm c}(M,z)
+
\gamma_{\rm E}\, y_{V}\,\hat{\mathcal{D}}_\nu\bar{I}^{\rm s, cyl}_\nu(M,z)
\\
\label{eq:gamma_E_Def}
\gamma_{\rm E}&=\frac{\left<I^{\rm s}_\nu(M,z)\right>_\tau}{\bar{I}^{\rm s, cyl}_\nu(M,z)}\equiv
\frac{\pi R^2\int\int
\frac{j_{\nu}^{\rm s}(r, M,z)\,P_{\rm e}(r')}{4\pi|\vek{r}-\vek{r}'|^2}\,{\rm d}V \,{\rm d}V'}
{\int j_{\nu}^{\rm s}(r, M,z)\,{\rm d}V
\int P_{\rm e}(r')\,{\rm d}V'},
\end{align}
\end{subequations}
where we introduced the energy-exchange form factor, $\gamma_{\rm E}$, for intracluster scattering of the halo. It can be computed once the pressure profile and emissivity profiles are defined. Note that frequency-dependence drops out of $\gamma_{\rm E}$ per definition. 
Also, as defined $\gamma_{\rm E}$ is expected to be of order unity and a weak functions of the mass and redshift. In Appendix~\ref{app:formfac} we give some details about how to estimate them. For our computations we shall use $\gamma_{\rm E}\approx 1.3$.

For the cosmological application one now has to replace $I_{\nu}^{\rm c}(M,z)$ and $\bar{I}^{\rm s, cyl}_\nu(M,z)$ with the related halo model luminosities. For the scattering $y$-parameter one can make a similar argument (see Appendix~\ref{app:y_cos}). Overall this lead to
\begin{subequations}
\label{eq:replacements}
\begin{align}
I_{\nu}^{\rm c}(M,z)&\rightarrow
\frac{1}{4\pi\,\chi^2\,a^3 }\,\frac{a L^{\rm c}_\nu(M,z)}{4\pi}
\\
\bar{I}^{\rm s, cyl}_\nu(M,z)&\rightarrow
\frac{1}{4\pi\,\chi^2\,a^3 }\,\frac{a L^{\rm s}_\nu(M,z)}{4\pi}
\\
y_V&\rightarrow
\frac{1}{4\pi\,d^2_A\,a^3}\,\frac{\sigma_{\rm T}}{m_{\rm e} c^2}\,\int P_{\rm e}(r)\,{\rm d}V
=\frac{y(M,z)}{4\pi\,a^3}.
\end{align}
\end{subequations}
where $y(M,z)$ is given by Eq.~\eqref{eq:y_M_def}. The total CIB contribution escaping from the halo then is
\begin{align}
\label{eq:I_tot_escape}
a^3\bar{I}_{\nu}^{\rm h}&\approx
\frac{1}{4\pi\,\chi^2} 
\frac{a L^{\rm h}_\nu(M,z)}{4\pi}
\nonumber\\
&\qquad
+\frac{y(M,z)}{4\pi} \left[
\frac{1}{2}
\hat{\mathcal{D}}_\nu \frac{a L^{\rm c}_\nu(M,z)}{4\pi}
+
\gamma_{\rm E}\, \hat{\mathcal{D}}_\nu \frac{a L^{\rm s}_\nu(M,z)}{4\pi}
\right].
\end{align}
The first term is the single-halo contribution to the unscattered average CIB, while the other two account for the distortion part caused by intracluster scattering. For an individual halo, these two contributions are usually much smaller than the ambient CIB flux and associated intercluster scattering of light from the other halos. However, once added up over all halos, it contributes at a similar level.

Putting everything together, the total intracluster contribution to the scattered CIB today is then
\begin{align}
\Delta I^{\rm intra}_{\nu, 0}
&\approx \int^\infty_{z}\frac{c\,{\rm d}z'}{H'}
\int {\rm d}M\,\frac{{\rm d}N(M,z')}{{\rm d}M}\,
y(M,z') \\ \nonumber
&\qquad\quad\times\left[
\frac{1}{2}
\frac{a' L^{\rm c}(M,z')}{4\pi}
+
\gamma_{\rm E}
\frac{a' L^{\rm s}(M,z')}{4\pi}\,
\right]\Bigg\}
\,\Delta \Theta_{\rm sc}(\nu/a', z')
\end{align}
after integrating over the halo mass function. Here it is noteworthy that the $y$-parameter is evaluated at the same redshift as the emission. This means, the scattered intracluster light will have a different SED than the scattered intercluster light, which did suffer redshifting first. 

Although our derivation neglected relativistic corrections, we can again take them into account by computing $\Delta \Theta_{\rm sc}(\nu/a', z')$ using the scattering kernel method. The corrections will behave very similar to those to the intercluster signal such that we do not give a more detailed discussion here.

\subsection{CIB distortion due to intracluster scattering}

\begin{figure}
\centering 
\includegraphics[width=\columnwidth]{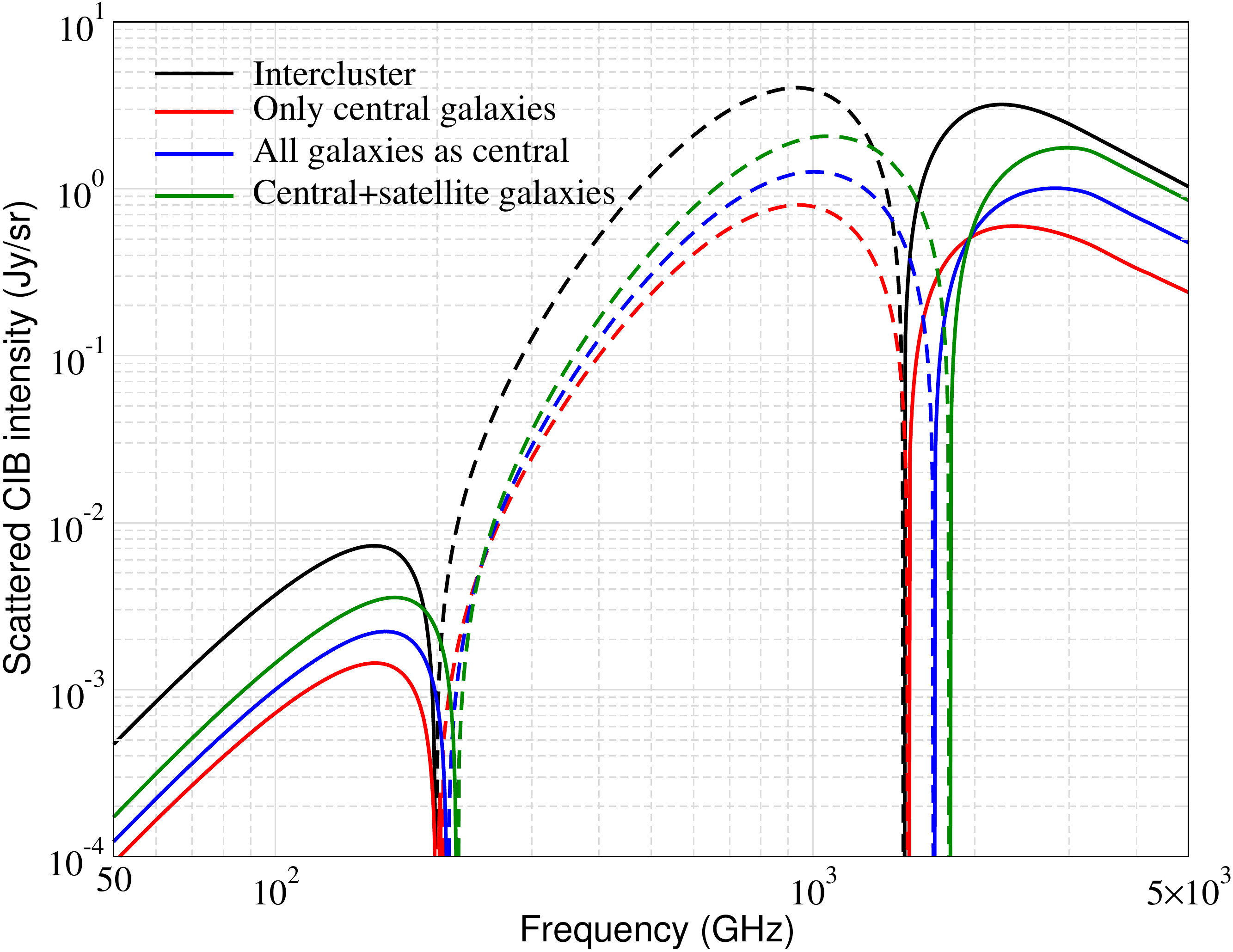}
\caption{Intensity of distortions from intercluster (black), only central galaxies (red), treating all galaxies as central (blue) and treating central and satellite galaxies separately with $\gamma_E=1.3$ . The solid line is the positive part and dashed line is the negative part of intensity.  }
\label{fig:total_distortion}
\end{figure}


\begin{figure}
\centering 
\includegraphics[width=\columnwidth]{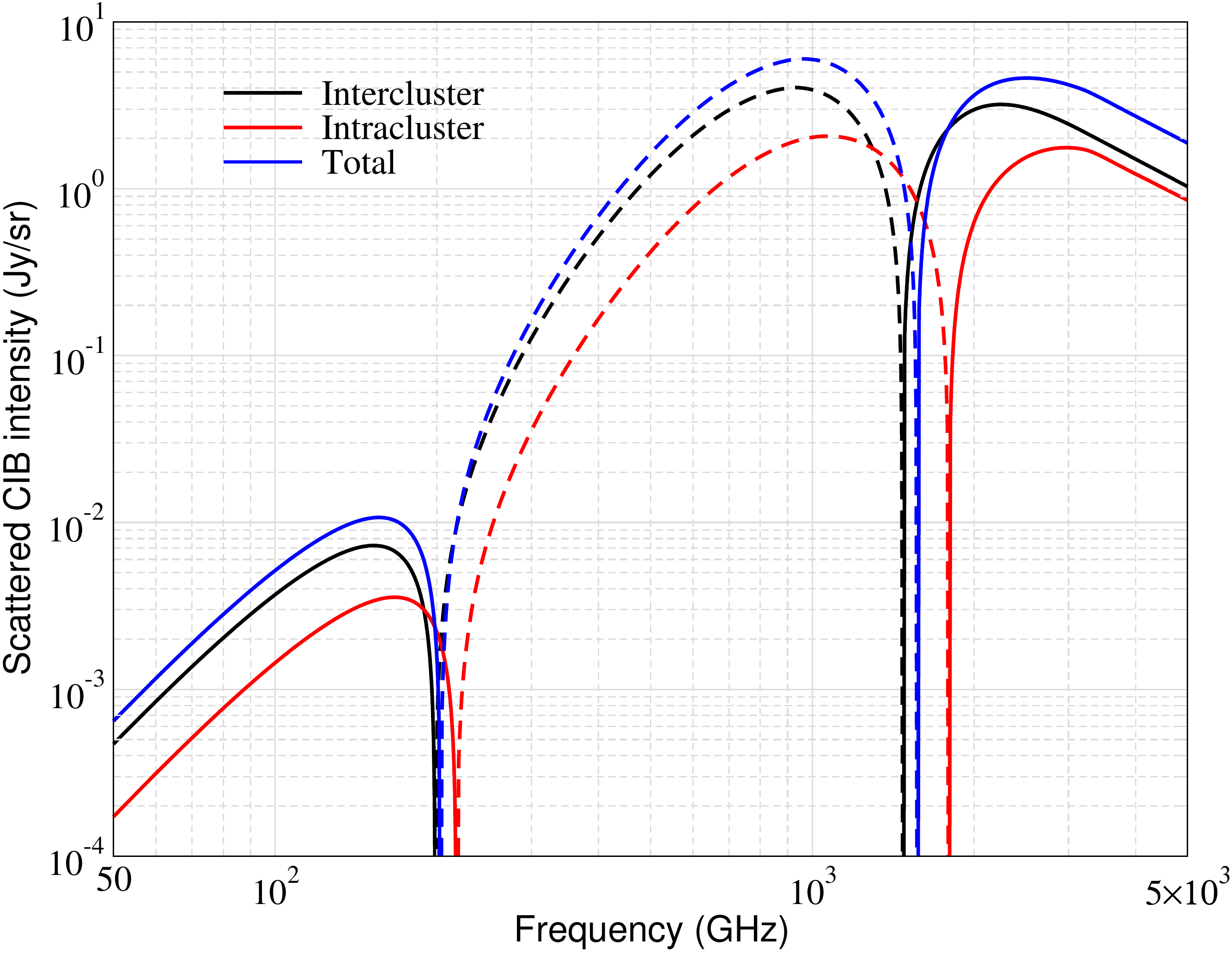}
\caption{Intensity of global averaged intercluster, intracluster and total distortions.  }
\label{fig:total_distortion1}
\end{figure}


\begin{figure}
\centering 
\includegraphics[width=\columnwidth]{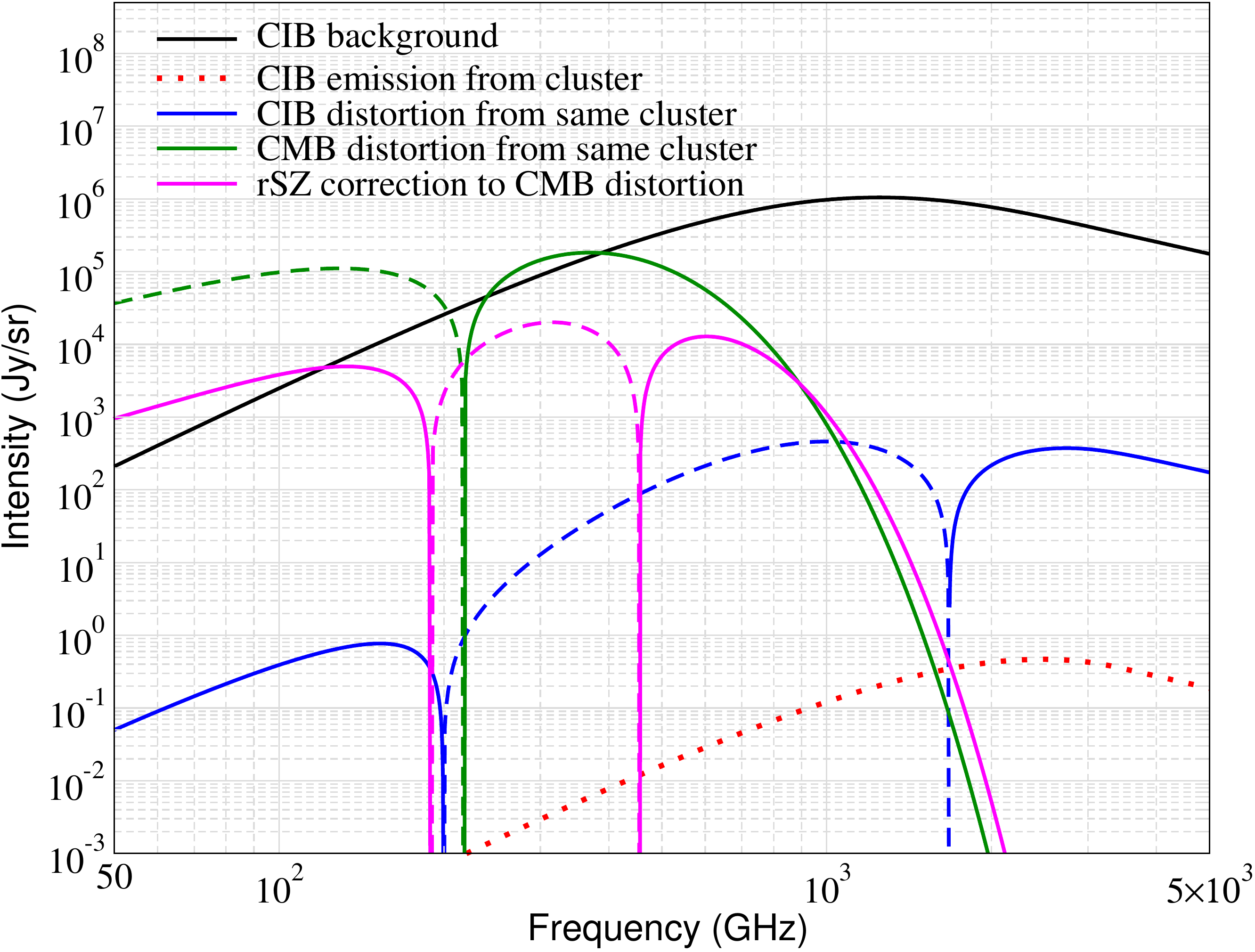}
\caption{Intercluster CIB distortion (including relativistic corrections) and CIB radiation from a cluster with mass $5\times 10^{14}M_{\odot}$ located at $z=0.1$ as seen today. The $y$-parameter for this cluster is $\approx 10^{-4}$ with temperature $\approx 6$ keV. We show the corresponding CMB distortion signal along with rSZ corrections. The CIB background, as seen today, is shown for reference.  }
\label{fig:cluster_distortion}
\end{figure}


\begin{figure}
\centering 
\includegraphics[width=\columnwidth]{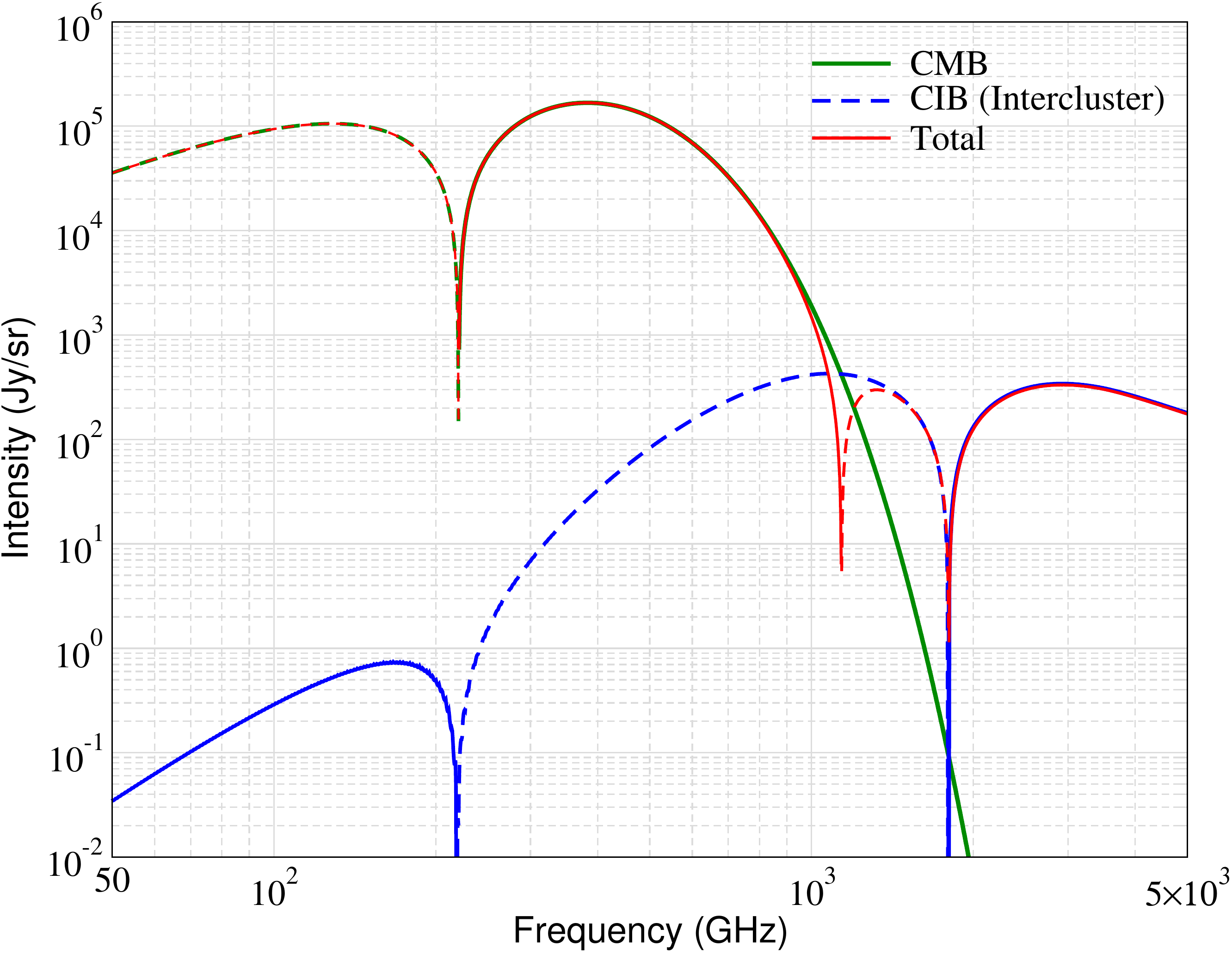}
\caption{CMB distortion (including rSZ corrections), CIB (intercluster) distortion and the sum of two for a galaxy cluster with parameters as those used in Fig.~\ref{fig:cluster_distortion}.}
\label{fig:cluster_distortion1}
\end{figure}


\begin{figure}
\centering 
\includegraphics[width=\columnwidth]{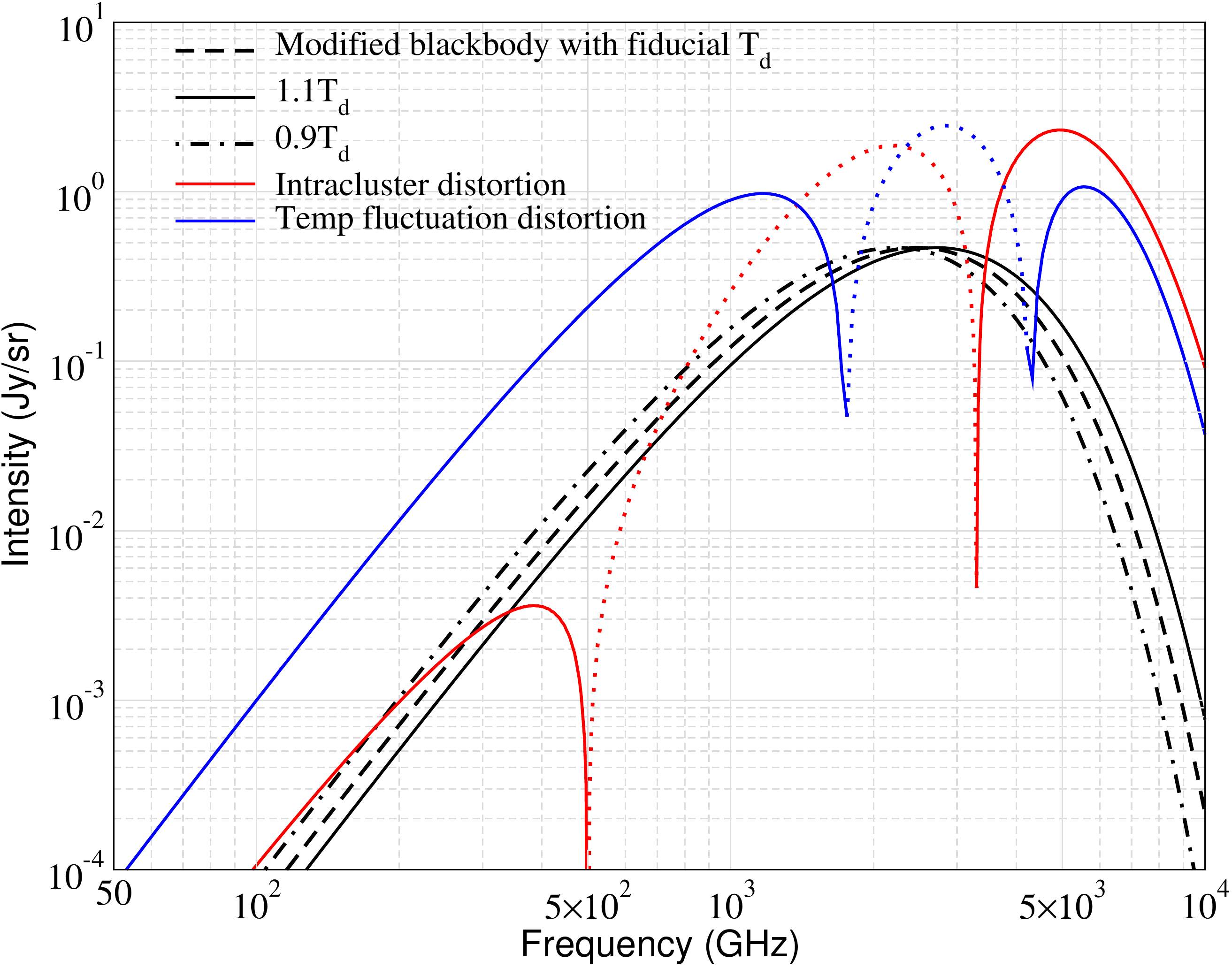}
\caption{Spectrum of IR radiation and its variation with temperature assuming it to be a modified blackbody spectrum. The parameters of the galaxy cluster are the same as in Fig.~\ref{fig:cluster_distortion}. We illustrate the intracluster scattering distortion setting the nomalization of the spectrum to $y_{\rm sc}=1$. Similarly, the distortion due to temperature variance inside the cluster is illustrated for an effective $y$-parameter, $y_{\rm var}=\frac{1}{2}(\Delta T_{\rm d}/T_{\rm d})^2=1$, to showcase the spectral properties. For comparison, we also give the modified blackbody spectra for three dust temperatures.}
\label{fig:temp_fluctuation1}
\end{figure}


In Fig. \ref{fig:total_distortion}, we plot the global averaged CIB distortions due to intracluster scattering and compare it with CIB distortions from intercluster scattering. As we have already stated in previous section, the contribution from intracluster scattering is of the similar order as that of intercluster scattering. As expected, the intracluster signal depends upon the galaxy distribution inside an halo which in our calculation is captured by the form factor $\gamma_E$. The intracluster signal has null points at higher frequency compared to intercluster signal with $\nu_{\rm low}\approx 218$ GHz and $\nu_{\rm high}=1800$ GHz. Most of the CIB distortion signals arise at $z\lesssim 1$, when heavier dark matter halos form. While the rest frame radiation spectrum from galaxies has a peak at $\approx$ few thousands (similar to $z=5$ curve in upper panel of Fig. \ref{fig:CIB_spectrum}), irrespective of redshift, the background CIB signal has redshifted to lower frequency ($z=1$ curve of same figure). This shifts the null point of intracluster contribution to higher frequency. The location of null point is sensitive to distribution of satellite galaxies. For dark matter halos with $M\lesssim 10^{13} M_{\odot}$, which forms at earlier redshifts, central galaxy gives the dominant contribution. Once heavier halos which can host satellite galaxies form at $z\lesssim$ 1, the contribution of satellite galaxies become important. This increases the contribution of $z\lesssim 1$ part of signal compared to earlier redshifts. This fact makes the null point shifts to higher frequency when we take satellite galaxies into account (blue lines in Fig. \ref{fig:total_distortion}) as compared to the case when we ignore them (red lines). Even though we have divided the the total signal to intercluster and intracluster type, we are only going to see the total signal which we show in Fig. \ref{fig:total_distortion1}. The total signal has null points at $\nu_{\rm low}\approx 203$ GHz and $\nu_{\rm high}=1570$ GHz. The result will be sensitive to the intracluster scattering form factor and details of the halo profile, offering a possibility to directly measure this parameter.

In Fig.~\ref{fig:cluster_distortion}, we plot the distortion signal expected from a single cluster located at $z=0.1$ with $M=5\times 10^{14}M_{\odot}$. The $y$-parameter of this cluster is  $\approx 10^{-4}$. We show the CIB distortion observed today from scattering of CIB background photons (i.e. intercluster signal) and direct CIB emission from the galaxies inside the cluster. The intensity of direct CIB emission is given by Eq.~\eqref{eq:Iav_initial} with $D\approx 390$~Mpc. The intracluster scattering signal is expected to be $\simeq 10^{-4}$ of direct CIB emission and thus is unimportant. The direct CIB emission also seems unimportant for detecting the CIB distortion from a single object (blue versus red lines). 

In Fig.~\ref{fig:cluster_distortion1}, we present the sum of the CMB and CIB (intercluster) distortion. At  $\nu\gtrsim 10^3$ GHz, the CIB distortion drastically changes the total signal that is expected in the direction of a galaxy cluster with respect to just the CMB distortion. We see that there are three null frequencies due to the superposition of both CMB and CIB distortion. From low to high frequencies, this leads to a characteristic sink-source-sink-source appearance of the cluster relative to the background.
At $\nu \lesssim 10~{\rm GHz}$ (not shown here), due to the radio-SZ effect we would expect the cluster to again become a source with respect to the background. A combination of measurements in all these regimes can thus in principle probe the evolution and origin of the ambient cluster radiation field.

By performing differential measurements on- and off-cluster, the sum of the direct CIB emission and intercluster cCIB will be important. For comparison, in Fig.~\ref{fig:cluster_distortion} we also show the tSZ signal and separate rSZ correction for the same cluster. As we can see, the intercluster scattering CIB signal is typically one order of magnitude lower than the rSZ signal. However, for high precision, especially in the Wien-tail, the cCIB signal should be modelled carefully (Fig.~\ref{fig:cluster_distortion1}). We leave a more detailed assessment to future work.

\subsection{Distortion due to dust temperature fluctuations}
\label{sec:T_variance}
Until now, we have assumed that individual galaxy clusters have a well defined SED temperature $T_{\rm d}(z)$. However, in addition to variation with redshift the CIB temperature can have fluctuations inside individual cluster.\footnote{We thank Eiichiro Komatsu for inquiring about this.} The mixture of SEDs with different temperatures will lead to a distortion on top of the average infrared spectrum, an effect that can again be modelled using temperature moments \citep{Chluba2017foregrounds}. This signal has nothing to do with scattering of photons with the electrons inside the cluster, but leads to
a $y$-type distortion with an effective $y$-parameter, $y_{\rm var}=\frac{1}{2}(\Delta T_{\rm d}/T_{\rm d})^2$.

For a CIB spectrum with occupation number $n_{\nu}$, keeping terms up to second order in $\Delta T_{\rm d}/T_{\rm d}$, we can write,
\begin{equation}
    \langle n_{\nu}(T_{\rm d}+\Delta T_{\rm d})\rangle_{\rm cl}=n_{\nu}(T_{\rm d})+ \frac{1}{2}\left<\left(\frac{\Delta T_{\rm d}}{T_{\rm d}}\right)^2\right>_{\rm cl}\, T_{\rm d}^2\frac{\partial^2 n_{\nu}}{\partial T_{\rm d}^2},
\end{equation}
where the linear term drops out after the average over the cluster around the mean dust temperature, $T_{\rm d}(z)$, is carried out. For the spectrum given by Eq.~\eqref{eq:define_theta}, one encounters a discontinuity in the second temperature derivative at $\nu=\nu_0$.\footnote{Indeed when using the Kompaneets equation to compute the cCIB signal, one also finds this, a conceptual problem that is avoided in our scattering kernel treatment.} To make the main points, we therefore assume the spectrum to be a given by the expression for $\nu\leq \nu_0$ at all frequencies. This simplification should not affect our conclusions once physical CIB spectra are used.

For a modified blackbody spectrum, Eq.~\eqref{eq:define_theta}, the temperature derivative acts on the terms $\Theta_\nu\propto (h\nu/\kB T_{\rm d})^{\beta+3}/(\expf{h\nu/\kB T_{\rm d}}-1)$, since $h\nu_0/\kB T_{\rm d}={\rm const}$ according to Eq.~\eqref{eq:x0_def}. The second temperature derivative of $n_\nu \propto \Theta_\nu/\nu^3$ can then be obtained as
\begin{equation}
\label{eq:define_sec_der_n}
\frac{T_{\rm d}^2}{n_{\nu}}\frac{\partial^2 n_{\nu}}{\partial T_{\rm d}^2}
=Y(x)-2(2+\beta) \,G(x)+(3+\beta)(4+\beta)
\end{equation}
Comparing to the corresponding scattering distortion in Eq.~\eqref{eq:define_theta_sc} we can see that the two spectra differ, and hence can in principle be distinguished using multi-frequency observation. 
This point is illustrated in Fig.~\ref{fig:temp_fluctuation1}. The variance distortion amplitude is determined by $y_{\rm var}=\frac{1}{2}\langle(\Delta T_{\rm d}/T_{\rm d})^2\rangle_{\rm cl}$, as already anticipated. Since we are mainly interested in differences of the spectral shapes, in Fig.~\ref{fig:temp_fluctuation1} we set both the electron scattering and the effective variance $y$-parameters to unity.
%

For an intracluster scattering signal with $y_{\rm sc}\simeq 10^{-4}$, the signal is already overwhelmed by temperature fluctuation with amplitude $\Delta T_{\rm d}/T_{\rm d}\approx 0.01$. However, by determining the scattering $y$-parameter with the CMB SZ observable, one can in principle differentiate the intracluster scattering signal and distortions due CIB temperature fluctuations.

We note that part of the variance and scattering signals can be fully absorbed by redefining the dust pivot temperature, since these are spectrally degenerate with a simple change of $T_{\rm d}$. To understand this, let us consider the total single cluster distortion,
\begin{align}
\label{eq:total_theta_sc_var}
\frac{\Delta \Theta_\nu}{\Theta_\nu}&=y_{\rm sc}\frac{\Delta \Theta_\nu}{\Theta_\nu}\Bigg|_{\rm sc}+
y_{\rm var}\frac{\Delta \Theta_\nu}{\Theta_\nu}\Bigg|_{\rm var}
\nonumber 
\\
&\approx 
(y_{\rm sc}+y_{\rm var})\,Y(x)
-2 \left[ y_{\rm sc} \beta + y_{\rm var}(2+\beta) \right] G(x)
\nonumber 
\\
&\qquad +\left[y_{\rm sc}\beta+y_{\rm var}(4+\beta)\right] (3+\beta).
\end{align}
The CIB temperature shift a first order in $\Delta T_{\rm d}/T_{\rm d}\ll 1$ has the simple spectrum 
$\frac{\Delta \Theta_\nu}{\Theta_\nu}=
\frac{T_{\rm d}}{n_{\nu}}\frac{\partial n_{\nu}}{\partial T_{\rm d}}=G(x)-(3+\beta)$. Hence, by lowering the pivot temperature to 
%
\begin{align}
T'_{\rm d}=T_{\rm d}\left\{1-2 \left[ y_{\rm sc} \beta + y_{\rm var}(2+\beta) \right]\right\}
\end{align}
we can fully absorb the term $\propto G(x)$.
Once redefining the CIB temperature to this value we find the total distortion
\begin{align}
\label{eq:total_theta_sc_var}
\frac{\Delta \Theta_\nu}{\Theta'_\nu}
&\approx 
(y_{\rm sc}+y_{\rm var})\,\left[Y(x)-\beta(3+\beta)
\right]
\end{align}
with respect to $\Theta'_\nu=\Theta_\nu(T'_{\rm d})$. In reality, this expression fully describes the new information that is added by scattering and temperature variance effects, since all other terms in the signal spectra can be modelled to high precision using the unscattered signal parameterization. The distortion then only has {\it one} null frequency at $x\simeq 5.468$. However, when performing differential measurements with respect to the background, the sink-source-sink-source structure will reappear, as explained above. More details about the observational aspects are, however, left to a future publication.


\section{Conclusions}
\label{sec:conclusions}
In this paper, we showcase the importance of intracluster scattering to the globally averaged CIB distortion signal which was recently considered in \cite{SHB2022}. Hot electrons inside dark matter halos can boost the CIB photons just like the CMB photons which distorts the CIB background. As opposed to CMB, which is truly a background radiation, CIB background is formed via emission of photons from galaxies inside the dark matter halos. These photons can be scattered by the electrons inside the parent halo or another halo after escaping from the parent halo. We term them as intracluster and intercluster distortions, respectively. The authors in \cite{SHB2022} considered only the intercluster contribution. We compute the intracluster contribution and show it to be equally important for the sky-averaged CIB distortion signal. 

The intercluster signal is sensitive to total luminosity of CIB photons from an individual halo while the intracluster signal is sensitive to the distribution of galaxies inside the halo too. To model this dependence, we have introduce a halo form factor which captures the information of satellite galaxy distribution inside a halo. The satellite galaxies have the effect of shifting the null points of total CIB distortion signal to higher frequency. Therefore, the measurement of CIB distortion signal can be used to infer the galaxy distribution inside the halo. Given that there is large uncertainty in the value of the form factor, further studies based on hydro simulations are needed. The findings could then potentially be confirmed with future CMB and far-infrared measurements. However, the CIB is expected to be lumpy which can make a detection difficult for individual cluster. The lumpy structure gives rise to CIB anisotropy which at cluster scale can be of the order of $10^{-2}-10^{-3}$Jy$^2$/sr \citep[see Fig. 8 of ][]{SHB2022}. Other emission signals such as CO lines \citep{MD2014} may further contaminate signal at $\lesssim 200$ GHz.  

We go beyond the assumption of non-relativistic Compton scattering between the hot electrons and the CIB photons which the authors in \cite{SHB2022} use. We take into account the relativistic corrections as the electrons inside the heaviest halos have temperature $\gtrsim$ 1 keV. These corrections are not important for sky-averaged distortions but are important for signal from an individual cluster which can have temperature $\gtrsim$ 5 keV.
This has the effect of shifting the null points of CIB distortion signal just as in case of CMB. As the CIB background is formed once structures form, it has a complicated redshift dependence as opposed to the CMB. This makes the shape of CIB distortion from an individual object redshift dependent, an effect that is not present for the tSZ. Therefore, one can in principle obtain a measure of the cluster gas temperature and its redshift by measuring both the CMB and CIB distortions. Also the CIB distortions has two null points as opposed to one for the CMB. These nulls move in opposite direction as we increase the temperature of gas which can be used to break the degeneracy of object's temperature and it's location. However, one needs to account for the direct CIB emission from individual cluster to detect the CIB distortion. 

In our computations we omitted the possible quadrupolar anisotropy in the average CIB at each location. This will lead to another correction to the intercluster cCIB that could affect the prediction. Turning this around, with cCIB measurements one could potentially constrain this correction. However, a more detailed discussion is beyond the scope of the work.

We also briefly discuss the corrections introduced by CIB temperature variance within individual clusters, highlighting that this signal can swamp the scattering signal even for very small temperature variance (see Sect.~\ref{sec:T_variance}). However, by combining multi-frequency measurements one can in principle determine the electron scattering $y$-parameter separately, and thereby learn about the intracluster temperature dispersion terms. A more detailed analysis is again beyond the scope of this paper.

We finally mention that a similar intracluster distortion contribution is expected in case of other radiation backgrounds that are created from within a halo. Radio background has recently been motivated to explain the detection of radio excess \citep{Fixsen2011excess,Edges2018}. However, these results are yet to be established and there is a lack of theoretical understanding of formation of such radio background as well.  All the ideas presented in this paper can be directly applied to the modeling of this backgrounds as well, once a model for the halo radio emission is introduced.

\vspace{-3mm}
{\small
\section*{Acknowledgments}

We thank Boris Bolliet, J. Colin Hill and Alina Sabyr for discussions related to their work.
We also thank Eiichiro Komatsu for his comments and inquiring about the CIB temperature variance effects.}
This work was supported by the ERC Consolidator Grant {\it CMBSPEC} (No.~725456).
JC was furthermore supported by the Royal Society as a Royal Society University Research Fellow at the University of Manchester, UK (No.~URF/R/191023).

\section{Data availability}
The data underlying in this article are available in this article and can further be made available on request.

{
\vspace{-3mm}
\bibliographystyle{mn2e}
\bibliography{Lit}

\begin{thebibliography}{46}
\expandafter\ifx\csname natexlab\endcsname\relax\def\natexlab#1{#1}\fi

\bibitem[{{Acharya} {et~al}\mbox{.}(2021){Acharya}, {Chluba}, \&
  {Sarkar}}]{Acharya2021FP}
{Acharya} S.~K., {Chluba} J., {Sarkar} A., 2021, \mnras

\bibitem[{{Battaglia} {et~al}\mbox{.}(2012){Battaglia}, {Bond}, {Pfrommer}, \&
  {Sievers}}]{BBPS2012}
{Battaglia} N., {Bond} J.~R., {Pfrommer} C., {Sievers} J.~L., 2012, \apj, 758,
  75

\bibitem[{{Birkinshaw}(1999)}]{Birkinshaw1999}
{Birkinshaw} M., 1999, Phys.~Rep, 310, 97

\bibitem[{{Bleem} {et~al}\mbox{.}(2015){Bleem}, {Stalder}, {de Haan}, {Aird},
  {Allen}, {Applegate}, {Ashby}, {Bautz}, {Bayliss}, {Benson}, {Bocquet},
  {Brodwin}, {Carlstrom}, {Chang}, {Chiu}, {Cho}, {Clocchiatti}, {Crawford},
  {Crites}, {Desai}, {Dietrich}, {Dobbs}, {Foley}, {Forman}, {George},
  {Gladders}, {Gonzalez}, {Halverson}, {Hennig}, {Hoekstra}, {Holder},
  {Holzapfel}, {Hrubes}, {Jones}, {Keisler}, {Knox}, {Lee}, {Leitch}, {Liu},
  {Lueker}, {Luong-Van}, {Mantz}, {Marrone}, {McDonald}, {McMahon}, {Meyer},
  {Mocanu}, {Mohr}, {Murray}, {Padin}, {Pryke}, {Reichardt}, {Rest}, {Ruel},
  {Ruhl}, {Saliwanchik}, {Saro}, {Sayre}, {Schaffer}, {Schrabback},
  {Shirokoff}, {Song}, {Spieler}, {Stanford}, {Staniszewski}, {Stark}, {Story},
  {Stubbs}, {Vanderlinde}, {Vieira}, {Vikhlinin}, {Williamson}, {Zahn}, \&
  {Zenteno}}]{Bleem2015}
{Bleem} L.~E. {et~al.}, 2015, \apjs, 216, 27

\bibitem[{{Bowman} {et~al}\mbox{.}(2018){Bowman}, {Rogers}, {Monsalve},
  {Mozdzen}, \& {Mahesh}}]{Edges2018}
{Bowman} J.~D., {Rogers} A. E.~E., {Monsalve} R.~A., {Mozdzen} T.~J., {Mahesh}
  N., 2018, \nat, 555, 67

\bibitem[{{Challinor} \& {Lasenby}(1998)}]{Challinor1998}
{Challinor} A., {Lasenby} A., 1998, \apj, 499, 1

\bibitem[{{Chluba} \& {Dai}(2014)}]{Chluba2014mSZ}
{Chluba} J., {Dai} L., 2014, \mnras, 438, 1324

\bibitem[{{Chluba} {et~al}\mbox{.}(2014){Chluba}, {Dai}, \&
  {Kamionkowski}}]{CDK2014}
{Chluba} J., {Dai} L., {Kamionkowski} M., 2014, \mnras, 437, 67

\bibitem[{{Chluba} {et~al}\mbox{.}(2017){Chluba}, {Hill}, \&
  {Abitbol}}]{Chluba2017foregrounds}
{Chluba} J., {Hill} J.~C., {Abitbol} M.~H., 2017, \mnras, 472, 1195

\bibitem[{{Chluba} {et~al}\mbox{.}(2012){Chluba}, {Nagai}, {Sazonov}, \&
  {Nelson}}]{ChlubaSZpack}
{Chluba} J., {Nagai} D., {Sazonov} S., {Nelson} K., 2012, \mnras, 426, 510

\bibitem[{{Chluba} \& {Sunyaev}(2009)}]{CS2009}
{Chluba} J., {Sunyaev} R.~A., 2009, \aap, 496, 619

\bibitem[{{Dowell} \& {Taylor}(2018)}]{DT2018}
{Dowell} J., {Taylor} G.~B., 2018, \apjl, 858, L9

\bibitem[{{Duffy} {et~al}\mbox{.}(2008){Duffy}, {Schaye}, {Kay}, \& {Dalla
  Vecchia}}]{DSKD2008}
{Duffy} A.~R., {Schaye} J., {Kay} S.~T., {Dalla Vecchia} C., 2008, \mnras, 390,
  L64

\bibitem[{{Erler} {et~al}\mbox{.}(2018){Erler}, {Basu}, {Chluba}, \&
  {Bertoldi}}]{Erler2017}
{Erler} J., {Basu} K., {Chluba} J., {Bertoldi} F., 2018, \mnras, 476, 3360

\bibitem[{{Fabbri}(1981)}]{Fabbri1981}
{Fabbri} R., 1981, \apss, 77, 529

\bibitem[{{Fixsen} {et~al}\mbox{.}(2011){Fixsen}, {Kogut}, {Levin}, {Limon},
  {Lubin}, {Mirel}, {Seiffert}, {Singal}, {Wollack}, {Villela}, \&
  {Wuensche}}]{Fixsen2011excess}
{Fixsen} D.~J. {et~al.}, 2011, \apj, 734, 5

\bibitem[{{Hill} {et~al}\mbox{.}(2015){Hill}, {Battaglia}, {Chluba}, {Ferraro},
  {Schaan}, \& {Spergel}}]{Hill2015}
{Hill} J.~C., {Battaglia} N., {Chluba} J., {Ferraro} S., {Schaan} E., {Spergel}
  D.~N., 2015, Physical Review Letters, 115, 261301

\bibitem[{{Hilton} {et~al}\mbox{.}(2021){Hilton}, {Sif{\'o}n}, {Naess},
  {Madhavacheril}, {Oguri}, {Rozo}, {Rykoff}, {Abbott}, {Adhikari}, {Aguena},
  {Aiola}, {Allam}, {Amodeo}, {Amon}, {Annis}, {Ansarinejad}, {Aros-Bunster},
  {Austermann}, {Avila}, {Bacon}, {Battaglia}, {Beall}, {Becker}, {Bernstein},
  {Bertin}, {Bhandarkar}, {Bhargava}, {Bond}, {Brooks}, {Burke}, {Calabrese},
  {Carrasco Kind}, {Carretero}, {Choi}, {Choi}, {Conselice}, {da Costa},
  {Costanzi}, {Crichton}, {Crowley}, {D{\"u}nner}, {Denison}, {Devlin},
  {Dicker}, {Diehl}, {Dietrich}, {Doel}, {Duff}, {Duivenvoorden}, {Dunkley},
  {Everett}, {Ferraro}, {Ferrero}, {Fert{\'e}}, {Flaugher}, {Frieman},
  {Gallardo}, {Garc{\'\i}a-Bellido}, {Gaztanaga}, {Gerdes}, {Giles}, {Golec},
  {Gralla}, {Grandis}, {Gruen}, {Gruendl}, {Gschwend}, {Gutierrez}, {Han},
  {Hartley}, {Hasselfield}, {Hill}, {Hilton}, {Hincks}, {Hinton}, {Ho},
  {Honscheid}, {Hoyle}, {Hubmayr}, {Huffenberger}, {Hughes}, {Jaelani}, {Jain},
  {James}, {Jeltema}, {Kent}, {Knowles}, {Koopman}, {Kuehn}, {Lahav}, {Lima},
  {Lin}, {Lokken}, {Loubser}, {MacCrann}, {Maia}, {Marriage}, {Martin},
  {McMahon}, {Melchior}, {Menanteau}, {Miquel}, {Miyatake}, {Moodley},
  {Morgan}, {Mroczkowski}, {Nati}, {Newburgh}, {Niemack}, {Nishizawa},
  {Ogando}, {Orlowski-Scherer}, {Page}, {Palmese}, {Partridge},
  {Paz-Chinch{\'o}n}, {Phakathi}, {Plazas}, {Robertson}, {Romer}, {Carnero
  Rosell}, {Salatino}, {Sanchez}, {Schaan}, {Schillaci}, {Sehgal}, {Serrano},
  {Shin}, {Simon}, {Smith}, {Soares-Santos}, {Spergel}, {Staggs}, {Storer},
  {Suchyta}, {Swanson}, {Tarle}, {Thomas}, {To}, {Trac}, {Ullom}, {Vale}, {Van
  Lanen}, {Vavagiakis}, {De Vicente}, {Wilkinson}, {Wollack}, {Xu}, \&
  {Zhang}}]{Hilton2021}
{Hilton} M. {et~al.}, 2021, \apjs, 253, 3

\bibitem[{{Holder} \& {Chluba}(2021)}]{HC2021}
{Holder} G., {Chluba} J., 2021, arXiv e-prints, arXiv:2110.08373

\bibitem[{{Hu} \& {Kravtsov}(2003)}]{HK2003}
{Hu} W., {Kravtsov} A.~V., 2003, \apj, 584, 702

\bibitem[{{Itoh} {et~al}\mbox{.}(1998){Itoh}, {Kohyama}, \& {Nozawa}}]{Itoh98}
{Itoh} N., {Kohyama} Y., {Nozawa} S., 1998, \apj, 502, 7

\bibitem[{Kompaneets(1956)}]{Kompa56}
Kompaneets A., 1956, Sov.Phys. JETP, 31, 876

\bibitem[{{Lee} {et~al}\mbox{.}(2022){Lee}, {Chluba}, \& {Holder}}]{LCH2022}
{Lee} E., {Chluba} J., {Holder} G.~P., 2022, \mnras

\bibitem[{{Lee} {et~al}\mbox{.}(2020){Lee}, {Chluba}, {Kay}, \&
  {Barnes}}]{Lee2020scalings}
{Lee} E., {Chluba} J., {Kay} S.~T., {Barnes} D.~J., 2020, \mnras, 493, 3274

\bibitem[{{Madau} \& {Dickinson}(2014)}]{MD2014}
{Madau} P., {Dickinson} M., 2014, \araa, 52, 415

\bibitem[{{McCarthy} \& {Madhavacheril}(2021)}]{MM2021}
{McCarthy} F., {Madhavacheril} M.~S., 2021, \prd, 103, 103515

\bibitem[{{Melin} {et~al}\mbox{.}(2018){Melin}, {Bartlett}, {Cai}, {De Zotti},
  {Delabrouille}, {Roman}, \& {Bonaldi}}]{Melin2018}
{Melin} J.~B., {Bartlett} J.~G., {Cai} Z.~Y., {De Zotti} G., {Delabrouille} J.,
  {Roman} M., {Bonaldi} A., 2018, \aap, 617, A75

\bibitem[{{Mroczkowski} {et~al}\mbox{.}(2019){Mroczkowski}, {Nagai}, {Basu},
  {Chluba}, {Sayers}, {Adam}, {Churazov}, {Crites}, {Di Mascolo}, {Eckert},
  {Macias-Perez}, {Mayet}, {Perotto}, {Pointecouteau}, {Romero}, {Ruppin},
  {Scannapieco}, \& {ZuHone}}]{SZreview2019}
{Mroczkowski} T. {et~al.}, 2019, Space Science Reviews, 215, 17

\bibitem[{{Planck Collaboration}(2014)}]{Planck_cib}
{Planck Collaboration}, 2014, \aap, 571, A30

\bibitem[{{Planck Collaboration} {et~al}\mbox{.}(2016){Planck Collaboration},
  {Aghanim}, {Arnaud}, {Ashdown}, {Aumont}, {Baccigalupi}, {Banday},
  {Barreiro}, {Bartlett}, {Bartolo}, \& et~al.}]{Planck2016ymap}
{Planck Collaboration} {et~al.}, 2016, \aap, 594, A22

\bibitem[{{Remazeilles} {et~al}\mbox{.}(2019){Remazeilles}, {Bolliet}, {Rotti},
  \& {Chluba}}]{RBRC2019}
{Remazeilles} M., {Bolliet} B., {Rotti} A., {Chluba} J., 2019, \mnras, 483,
  3459

\bibitem[{{Remazeilles} \& {Chluba}(2020)}]{RC2020}
{Remazeilles} M., {Chluba} J., 2020, \mnras, 494, 5734

\bibitem[{{Rephaeli}(1995)}]{Rephaeli1995}
{Rephaeli} Y., 1995, \apj, 445, 33

\bibitem[{{Rybicki} \& {dell'Antonio}(1994)}]{RD1994}
{Rybicki} G.~B., {dell'Antonio} I.~P., 1994, \apj, 427, 603

\bibitem[{{Rybicki} \& {Lightman}(1979)}]{Rybicki1979}
{Rybicki} G.~B., {Lightman} A.~P., 1979, {Radiative processes in astrophysics}.
  New York, Wiley-Interscience, 1979.~393 p.

\bibitem[{{Sabyr} {et~al}\mbox{.}(2022){Sabyr}, {Hill}, \& {Bolliet}}]{SHB2022}
{Sabyr} A., {Hill} J.~C., {Bolliet} B., 2022, arXiv e-prints, arXiv:2202.02275

\bibitem[{{Sarkar} {et~al}\mbox{.}(2019){Sarkar}, {Chluba}, \&
  {Lee}}]{CSpack2019}
{Sarkar} A., {Chluba} J., {Lee} E., 2019, \mnras, 490, 3705

\bibitem[{{Sazonov} \& {Sunyaev}(1998)}]{Sazonov1998}
{Sazonov} S.~Y., {Sunyaev} R.~A., 1998, \apj, 508, 1

\bibitem[{{Sazonov} \& {Sunyaev}(2000)}]{Sazonov2000}
{Sazonov} S.~Y., {Sunyaev} R.~A., 2000, \apj, 543, 28

\bibitem[{{Seljak}(2000)}]{S2000}
{Seljak} U., 2000, \mnras, 318, 203

\bibitem[{{Shang} {et~al}\mbox{.}(2012){Shang}, {Haiman}, {Knox}, \&
  {Oh}}]{SZKO2012}
{Shang} C., {Haiman} Z., {Knox} L., {Oh} S.~P., 2012, \mnras, 421, 2832

\bibitem[{{Sunyaev} \& {Zeldovich}(1980)}]{Sunyaev1980}
{Sunyaev} R.~A., {Zeldovich} I.~B., 1980, \mnras, 190, 413

\bibitem[{{Tinker} {et~al}\mbox{.}(2008){Tinker}, {Kravtsov}, {Klypin},
  {Abazajian}, {Warren}, {Yepes}, {Gottl{\"o}ber}, \& {Holz}}]{TKKAWYGH2008}
{Tinker} J., {Kravtsov} A.~V., {Klypin} A., {Abazajian} K., {Warren} M.,
  {Yepes} G., {Gottl{\"o}ber} S., {Holz} D.~E., 2008, \apj, 688, 709

\bibitem[{{Tinker} \& {Wetzel}(2010)}]{TW2010}
{Tinker} J.~L., {Wetzel} A.~R., 2010, \apj, 719, 88

\bibitem[{{Wright}(1979)}]{Wright1979}
{Wright} E.~L., 1979, \apj, 232, 348

\bibitem[{{Zeldovich} \& {Sunyaev}(1969)}]{Zeldovich1969}
{Zeldovich} Y.~B., {Sunyaev} R.~A., 1969, \apss, 4, 301

\end{thebibliography}
}

\begin{appendix}

\section{Derivation of the CIB solution in the expanding Universe}
\label{app:detailed_derive}
To obtain the solution for the average ambient CIB intensity at a given redshift, we go step by step through the derivation. We start with the general kinetic equation in the expanding (isotropic) Universe with a photon source term. We then clarify the connection of the halo luminosity to the required photon source.

\subsection{Kinetic equation for photon field in an expanding Universe}
We can write the kinetic equation for the evolution of the photon field in an isotropic medium as \citep[e.g.,][]{RD1994, CS2009},
\begin{equation}
    \frac{1}{c}\left[\frac{\partial N_{\nu}}{\partial t}\bigg|_{\nu}+2HN_{\nu}-H\nu\frac{\partial N_{\nu}}{\partial \nu}\bigg|_t\right]=C[N_{\nu}],
    \label{eq:kinetic_equation_nu}
\end{equation}
where $N_{\nu}=I_{\nu}/h\nu$ with $I_{\nu}$ being the physical intensity of the radiation field, $H(z)$ is the Hubble parameter and $C[N_{\nu}]$ is the collision term, which describes photon emission, absorption or interaction. 

To obtain the evolution of the background CIB, we simply assume that photons are emitted at some rate defined by the galaxy luminosity function. Overall, this means that we can specify an isotropic photon source term, $S(t, \nu)$, or emissivity. It is convenient to rewrite the kinetic equation in terms of $I_\nu$, which simply yields
\begin{equation}
    \frac{1}{c}\left[\frac{\partial I_{\nu}}{\partial t}\bigg|_{\nu}+3HI_{\nu}-H\nu\frac{\partial I_{\nu}}{\partial \nu}\bigg|_t\right]=S(t,\nu),
    \label{eq:kinetic_equation_nu_I}
\end{equation}
To solve this equation, we now perform a few transformations. Using the frequency variable, $x=\nu/(1+z)=a \nu$, we have
    $I_{\nu}=\frac{\rm{d}x}{\rm{d}\nu}I_x
    =a\,I_x$.
Inserting this into Eq.~\eqref{eq:kinetic_equation_nu_I}, we obtain
\begin{equation}
     \frac{1}{c}\left[\frac{\partial I_x}{\partial t}\bigg|_{\nu}+4H I_x-Hx\frac{\partial I_x}{\partial x}\bigg|_t\right]=\frac{S(t,x/a)}{a}.
    \label{eq:kinetic_equation_x}
\end{equation}
Using the total differential of $I_x$,
\begin{equation}
    {\rm d} I_x=\frac{\partial I_x}{\partial t}\bigg|_x {\rm d}t+\frac{\partial I_x}{\partial x}\bigg|_t {\rm d}x\rightarrow \frac{\partial I_x}{\partial t}\bigg|_\nu=\frac{\partial I_x}{\partial t}\bigg|_x+\frac{\partial I_x}{\partial x}\bigg|_t\,\frac{\partial x}{\partial t}\bigg|_x,
\end{equation}
with $\partial x/\partial t |_x=H x$,
and inserting into Eq.~\eqref{eq:kinetic_equation_x}, we find
\begin{equation}
     \frac{1}{c}\left[\frac{\partial I_x}{\partial t}\bigg|_{x}+4H I_x\right]=\frac{S(t,x/a)}{a}.
\end{equation}
To absorb the expansion term , we can transform to comoving quantities, $\tilde{I}_x=I_x/(1+z)^4\equiv a^3\,I_{\nu}$. In terms of $\tilde{I}_x$, we then find
\begin{equation}
\label{eq:evol_transformed}
    \frac{1}{c}\frac{\partial \tilde{I}_x}{\partial t}\bigg|_{x}=a^3 S(t, x/a).
\end{equation}
Rewriting things in terms of redshift, the solution then reads
\begin{equation}
    \tilde{I}_x(z)=\int^{z_{\rm max}}_z \frac{c\,{\rm d} z'}{H(z')}
    \,a'^4 S(z', x/a'),
\end{equation}
where we assume that at $z_{\rm max}$ we have $\tilde{I}_x=0$. Reverting back to physical coordinates we then have
\begin{equation}
\label{eq:sol_Inu}
    I_\nu(z)=\frac{1}{a^3}\int^{z_{\rm max}}_z  \,\frac{c \,{\rm d} z'}{H(z')}
    \,a'^4 S(z', \nu \,a/a').
\end{equation}
We highlight that $S(z, \nu)$ is the physical emissivity of the medium. The factor of $a^4$ makes the emissivity invariant under the expansion (see Appendix~\ref{sec:invariants}).
This means one simply has to provide the physical emissivity and integrate it in a comoving way and then afterwards divide by $a^3$ to obtain the solution at $z$. 
We now need to link this source term to the halo model quantities.

\vspace{-4mm}
\subsection{Connecting luminosity and emissivity}
In our computations, we will base the photon source term, $S(z, \nu)$, on the halo model. The halo model specifies the total CIB luminosity of each halo, $L^{\rm h}_\nu(M, z)$, as a function of halo mass $M$ and redshift $z$. The halo luminosity is defined as the physical luminosity of the halo when observed at $z$.
To obtain the photon source term, we need to link this to the emissivity of the halo, $j^{\rm h}_\nu(r, M,z)$, at a given location $\vek{r}$ inside the halo, which we assume to be spherically-symmetric. This links to the average intensity of the halo, which then links to the flux and finally the luminosity. 

\begin{figure}
\centering 
\includegraphics[width=\columnwidth]{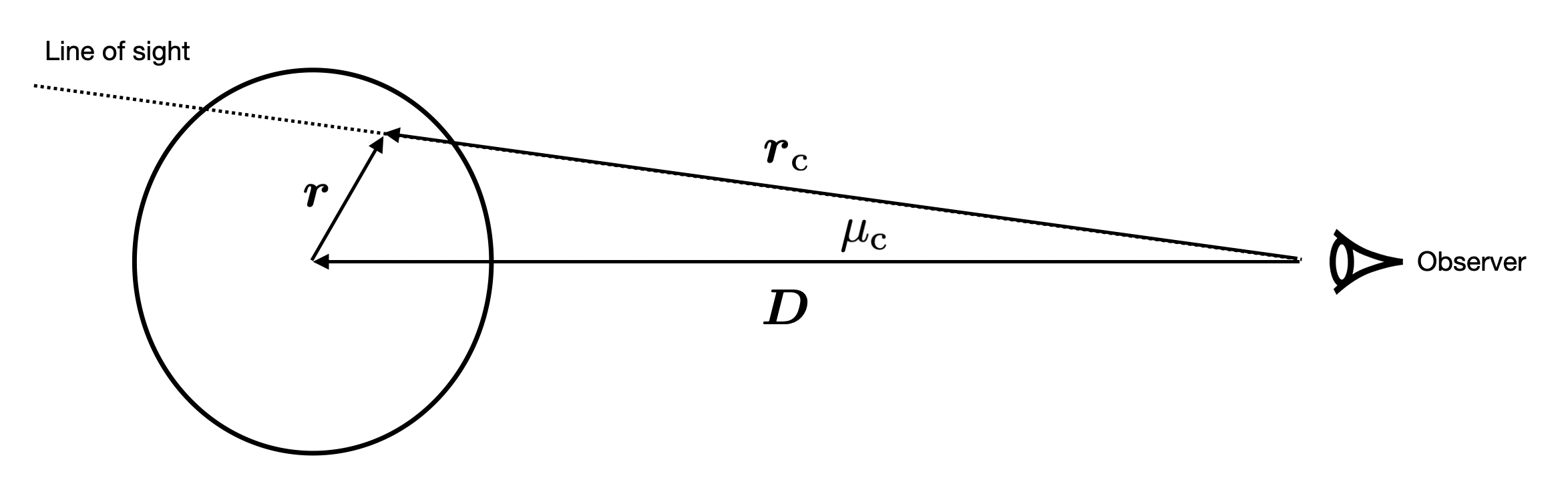}
\\
\caption{Computation of the average flux over a cluster.}
\label{fig:illustration_1}
\vspace{-3mm}
\end{figure}

Let us start with the flux integral over the halo. Assuming that the cluster is at a large distance, $D$, and centering the coordinate system on the observer, one has (see Fig.~\ref{fig:illustration_1})
\begin{align}
F_\nu &= \int I^{\rm h}_\nu(\theta_{\rm c}, \phi_{\rm c}, M,z) 
\cos\theta_{\rm c} \,{\rm d}\Omega_{\rm c}
\nonumber\\
&=
\int \int^\infty_0  j_{\nu}^{\rm h}(|\vek{D}-\vek{r}_{\rm c}|, M,z)\, {\rm d}r_{\rm c}
\cos\theta_{\rm c}\,{\rm d}\Omega_{\rm c}
\nonumber\\
&=
\int \int^\infty_0  \frac{j_{\nu}^{\rm h}(|\vek{D}-\vek{r}_{\rm c}|, M,z)}{r_{\rm c}^2}\, r_{\rm c}^2\,{\rm d}r_{\rm c}
\cos\theta_{\rm c}\,{\rm d}\Omega_{\rm c}
\nonumber\\
&\approx 
\frac{1}{D^2}\,\int j_{\nu}^{\rm h}(r, M,z)\,{\rm d}V,
\end{align}
where $(r_{\rm c}, \theta_{\rm c}, \phi_{\rm c})$ determines the locus of the emission point and $\vek{D}$ the cluster center. 
We used that the physical intensity along the line of sight is $I^{\rm h}_\nu(\theta_{\rm c}, \phi_{\rm c}, M,z)=\int^\infty_0  j_{\nu}^{\rm h}(r, M,z)\, {\rm d}r_{\rm c}$ with $r=|\vek{D}-\vek{r}_{\rm c}|$.
In the last line we used $\cos\theta_{\rm c}\approx 1$ and $r_{\rm c}\approx D$, which are good approximations once the integrations are performed relative to the cluster center, ${\rm d}V=r^2{\rm d}r\,{\rm d}\Omega$. The total CIB luminosity of the halo is then given by the integral over a spherical surface with radius $D$:
\begin{align}
\label{eq:L_j_relation_Fnu}
L^{\rm h}_\nu(M,z)=\int F_\nu\,{\rm d} A 
&= \int F_\nu \,D^2 \,{\rm d}\Omega
\approx 
4\pi\int j_{\nu}^{\rm h}(r, M,z)\,{\rm d}V.
\end{align}
Although here the argument was performed assuming a stationary flat space, it also translates to the expanding Universe. Using the invariance of the luminosity (see Appendix~\ref{sec:invariants}), in comoving coordinates one then has
\begin{align}
\label{eq:L_j_relation}
a L^{\rm h}_\nu(M,z)=4\pi\,a \int j_{\nu}^{\rm h}(r, M,z)\,{\rm d}V\equiv 4\pi\int \tilde{j}_{\nu}^{\rm h}(r, M,z)\,{\rm d}\tilde{V},
\end{align}
where it is understood that $\nu=x/a$ in terms of the comoving frequency. With the equations we can link the emissivity profile and cluster CIB luminosity function.

We still need to link the average CIB intensity at a given redshift to the luminosity function. To compute the average CIB intensity from a cluster at a fixed redshift, we need to first average over all lines of sight through the cluster. This is very similar to the flux integral but without the factor of $\cos\theta_{\rm c}$:
\begin{align}
\bar{I}^{\rm h}_\nu &= \int I^{\rm h}_\nu(\theta_{\rm c}, \phi_{\rm c}, M,z) 
\,\frac{{\rm d}\Omega_{\rm c}}{4\pi}
\nonumber\\
&=
\int \int^\infty_0  j_{\nu}^{\rm h}(|\vek{D}-\vek{r}_{\rm c}|, M,z)\, {\rm d}r_{\rm c}\,\frac{{\rm d}\Omega_{\rm c}}{4\pi}
\nonumber\\
&=
\int \int^\infty_0  \frac{j_{\nu}^{\rm h}(|\vek{D}-\vek{r}_{\rm c}|, M,z)}{r_{\rm c}^2}\, r_{\rm c}^2\,{\rm d}r_{\rm c}\,\frac{{\rm d}\Omega_{\rm c}}{4\pi}
\nonumber\\
&\approx 
\label{eq:Iav_initial}
\frac{1}{4\pi D^2}\,\int j_{\nu}^{\rm h}(r, M,z)\,{\rm d}V
= \frac{1}{4\pi D^2}\,\frac{L^{\rm h}_\nu(M,z)}{4\pi}.
\end{align}
To obtain the comoving intensity we then have 
\begin{align}
a^3 \bar{I}^{\rm h}_\nu 
&=\frac{1}{4\pi\,\chi^2}\,\frac{a L^{\rm h}_\nu(M,z)}{4\pi},
\end{align}
where we had to replace $D$ by the comoving distance, $\chi=\int c{\rm d}z'/H'$.

In every redshift shell we have a comoving number of halos
\begin{align}
{\rm d}N=\frac{{\rm d}^2 V}{{\rm d} z\,{\rm d}\Omega}\,\frac{{\rm d}N}{{\rm d}M}\,{\rm d}z\,{\rm d}\Omega\,{\rm d}M=\frac{c}{H}\,4\pi\,\chi^2\,\frac{{\rm d}N}{{\rm d}M}\,{\rm d}z
\end{align}
contributing to the total average signal. Here ${\rm d}^2 V/{\rm d} z\,{\rm d}\Omega$ is the comoving volume element per steradian and ${\rm d}N/{\rm d}M$ is the halo mass function, which determines the comoving number density of halos at various redshifts.
Putting things together, then yields
\begin{align}
a^3 I^{\rm CIB}_\nu(z)
&=\int^\infty_{z} \int \frac{{\rm d}^2 V}{{\rm d} z'\,{\rm d}\Omega}\,\frac{{\rm d}N}{{\rm d}M}\,{\rm d}z'\,{\rm d}\Omega\,{\rm d}M
\left(\frac{1}{4\pi\,\chi'^2}\,\frac{a' L^{\rm h}_{\nu'}(M,z')}{4\pi}\right)
\nonumber\\
&=\int^\infty_{z}\frac{c\,{\rm d}z'}{H'}
\int {\rm d}M\,\frac{{\rm d}N(M,z')}{{\rm d}M}\,
\frac{a' L^{\rm h}_{\nu'}(M,z')}{4\pi}.
\end{align}
with $\nu'=\nu \,a/a'$, where $\nu$ is fixed at the observer redshift $z$. By comparing with Eq.~\eqref{eq:sol_Inu} we then finally have
\begin{equation}
    a^4 S(z,\nu)=\int {\rm d}M\,\frac{{\rm d}N(M,z)}{{\rm d}M}\,\frac{a\,L^{\rm h}_\nu(M,z)}{4\pi}
    =\frac{a\,\bar{L}^{\rm h}_\nu(z)}{4\pi},
    \label{eq:averaged_source_term}
\end{equation}
where we introduced the mass-function-averaged luminosity $\bar{L}^{\rm h}_\nu(z)$. This yields the solution given in Eq.~\eqref{eq:sol_ICIB}.

\section{Invariants of radiation quantities in the expanding Universe}
\label{sec:invariants}
We already saw that in the expanding Universe $a^3 I_\nu={\rm const}$. With this we can also demonstrate several additional invariants. This can be readily seen when considering the definition of intensity \citep{Rybicki1979}
\begin{align}
{\rm d}E=I_\nu \,{\rm d}\nu\,{\rm d}A\,{\rm d}\Omega\,{\rm d}t.
\end{align}
In the expanding Universe, the energy differential $a{\rm d}E = a'{\rm d}E'$ is invariant. Since ${\rm d}A'/a'^2={\rm d}A/a^2$, we then have
\begin{align}
&a I_\nu \,{\rm d}\nu\,{\rm d}A\,{\rm d}\Omega\,{\rm d}t
\equiv a^3 I_\nu \,{\rm d}\nu\,{\rm d}\tilde{A}\,{\rm d}\Omega\,{\rm d}t
=
\tilde{I}_\nu \,{\rm d}\nu\,{\rm d}\tilde{A}\,{\rm d}\Omega\,{\rm d}t
\nonumber\\[1mm]
&\qquad\leftrightarrow
\tilde{I}_\nu \,{\rm d}\nu\,{\rm d}\tilde{A}\,{\rm d}\Omega\,{\rm d}t\equiv\tilde{I}'_{\nu'} \,{\rm d}\nu'\,{\rm d}\tilde{A}'\,{\rm d}\Omega'\,{\rm d}t'
\end{align}
where 'tilde' denotes comoving quantities. The last line then implies ${\rm d}\nu\,{\rm d}\Omega\,{\rm d}t$ is invariant. 

Since $a {\rm d}\nu$ is constant and ${\rm d}t/a$ is constant this also means ${\rm d}\Omega$ is invariant. 
Since flux is ${\rm d}F_\nu=I_{\nu} \cos\theta\,{\rm d}\Omega$, one then directly has $a^3 d F_\nu={\rm const}$. Similarly, for the emissivity, we use 
\begin{align}
{\rm d}E=j_\nu \,{\rm d}\nu\,{\rm d}V\,{\rm d}\Omega\,{\rm d}t
\end{align}
and thus $a^4 j_\nu={\rm const}$, where we used ${\rm d}V/a^3={\rm const}$. 
Finally, since the definition of luminosity is ${\rm d}L_\nu=I_{\nu} \cos\theta\,{\rm d}\Omega\,{\rm d}A$, it follows $a L_\nu={\rm const}$. For all bolometric quantities (i.e., integral over ${\rm d}\nu$) another factor of $a$ appears (e.g., $I=\int I_\nu\,{\rm d}\nu\rightarrow a^4 I={\rm const}$). These relations will be useful in multiple derivations presented here.

\vspace{-3mm}
\subsection{Transformation of the $y$-parameter}
\label{app:y_cos}
The transformation of the $y$-parameter deserves some extra attention. Independent of the frame, the line of sight number of scattering or total energy transfer have to be conserved. This means 
\begin{align}
\frac{{\rm d}y}{{\rm d}s}\,{\rm d}s=\theta_{\rm e}\,N_{\rm e}\,\sigma_{\rm T}\,{\rm d}s=\frac{1}{a^2} \, \theta_{\rm e}\,\tilde{N}_{\rm e}\,\sigma_{\rm T}\,{\rm d}\tilde{s}\equiv\frac{{\rm d}\tilde{y}}{{\rm d}\tilde{s}}\,{\rm d}\tilde{s}.
\end{align}
The comoving line of sight integral over the $y$-parameter in the expanding Universe then is
\begin{align}
\tilde{y}(\hat{\vek{\gamma}})&=\int \frac{{\rm d}\tilde{y}}{{\rm d}\tilde{s}}\,{\rm d}\tilde{s}=\frac{1}{a^2} \int \theta_{\rm e}\,\tilde{N}_{\rm e}\,\sigma_{\rm T}\,{\rm d}\tilde{s}.
\end{align}
The $y$-parameter per unit solid angle, towards the cluster then gives
\begin{align}
\bar{\tilde{y}}&=\int \tilde{y}(\hat{\vek{\gamma}})\,\frac{{\rm d}\Omega}{4\pi}=\frac{1}{4\pi\,a^2} \int \theta_{\rm e}\,\tilde{N}_{\rm e}\,\sigma_{\rm T}\,{\rm d}\tilde{s}\,\Omega
\nonumber\\
&=\frac{1}{4\pi\,\chi^2\,a^2}\,\frac{\sigma_{\rm T}}{m_{\rm e} c^2}\,\int \tilde{P}_{\rm e}(r)\,{\rm d}\tilde{V}
\nonumber\\
&=\frac{1}{4\pi}\,\frac{\sigma_{\rm T}\,}{m_{\rm e} c^2}\,\frac{4\pi\,\tilde{r}^3_\Delta(z)}{d_A^2(z)}\int P_{\rm e}(x)\,x^2\,{\rm d}x=\frac{y(M,z)}{4\pi\,a^3}.
\end{align}
The factor of $1/4\pi$ will disappear after we integrate over the solid angle for full sky to obtain the total signal.


\vspace{-3mm}
\section{Fit for $\mathcal{S}(z)$}
\label{app:fit_Lz}
For our fixed cosmology and halo parameters we find 
\begin{align}
\label{eq:Sfit}
\mathcal{S}(z)
&=\frac{c\,a\,\bar{L}^{\rm h}(z)}{4\pi\, H(z)}
\nonumber \\
&\approx\exp\Bigg(
2884 (1+z)^{2.6}\,[1 - 9.715 (1 + z) + 165.7 (1 + z)^2
\nonumber \\
&\quad + 89.2 (1 + z)^3  + 
 1.30 (1 + z)^4 + 0.066 (1 + z)^6]^{-1}
\Bigg)
\end{align}
to work to $1\%$ precision at $z\leq 6$. For numerical applications this provides a very useful benchmark. Note that this approximation includes both the sub-halo and central galaxy contributions.

\section{Fit for $\xi^*(z)$}
\label{app:fitxi}
For our fixed cosmology and halo parameters we find 
\begin{align}
\label{eq:xifit}
\xi^*(z)
&\approx\frac{2.544}
{1 + 0.6390 z - 0.1060 z^2 + 0.0060 z^3 + 0.00022 z^4}
\end{align}
to work to $1\%$ precision at $z\leq 5$. 
\section{The intracluster scattering form factors}
\label{app:formfac}
To compute the intracluster scattering form factors we have to fix the pressure and halo emissivity profiles. For this, only the shapes of the profiles are important. 
However, before going into details, let us consider a constant pressure and emissivity profile within a sphere of radius $R$. In this case, we have 
\begin{align}
\gamma^{\rm c}_{\rm E}=
\frac{\pi R^2\int\int
\frac{{\rm d}V \,{\rm d}V'}{4\pi|\vek{r}-\vek{r}'|^2}}
{\int {\rm d}V \int {\rm d} V'}=\frac{9}{16}.
\end{align}
where we used $\int\int
{\rm d}V \,{\rm d}V'/4\pi|\vek{r}-\vek{r}'|^2=\pi R^4$. In the more realistic cases we find a form factor more close to unity.

For the pressure, we follow \citet{BBPS2012}:
\begin{equation}
    P_{\rm e}(r)\propto (x/x_c)^{\gamma}\left[1+(x/x_c)^{\alpha}\right]^{-\beta}.
\end{equation}
where $x=r/r_{200\rm c}$, $\alpha=1$ and $\gamma=-0.3$, while $x_c$ and $\beta$ depend on the halo properties. The expression for $x_c$ and $\beta$ as a function of $M_{200\rm c}$ and $z$ can be found in \cite{BBPS2012} and read
\begin{subequations}
\begin{align}
x_c&=0.497\,\left[\frac{M_{200\rm c}}{10^{14}\,M_{\odot}}\right]^{-0.00865} (1 + z)^{0.731}
\\
\beta&=4.35\,\left[\frac{M_{200\rm c}}{10^{14}\,M_{\odot}}\right]^{0.0393} (1 + z)^{0.415};
\end{align}
\end{subequations}
For the emissivity profile we assume that is similar to NFW profile:
\begin{equation}
    j_\nu(r, M, z)\propto\frac{1}{(r/r_s)(1+r/r_s)^2},
\end{equation}
where $r_s$ is the characteristic scale of the profile. Since the SED is an overall factor below we drop the index $\nu$ for convenience. We convert $r_s$ to $r_{200\rm c}$ using the relation \citep{DSKD2008}
\begin{equation}
c_{200\rm c}=5.71\,\left[\frac{M_{200\rm c}}{2\times 10^{12}h^{-1} M_{\odot}}\right]^{-0.084}
(1+z)^{-0.47}
\end{equation}
for the concentration parameter. The two distance scales $r_s$ and $r_{200c}$ are related by, $c_{200c}=r_{200c}/r_s$ and $r/r_s=c_{200c}\,x$.

The multidimensional integrals in Eq.~\eqref{eq:gamma_E_Def} are quite hard to compute. We used various numerical methods and also studied the dependence on $M_{200\rm c}$ and $z$. We find that $\gamma_{\rm E}$ can indeed vary quite significantly (even by an order of magnitude) depending on the maximal radius that the integrals are carried out to. In addition, the core-excision affects the results based on the profiles that we used. For our applications, we will use a fiducial value $\gamma_{\rm E}=1.3$ and defer a more detailed investigation to future work.

\end{appendix}

\end{document}